\documentclass{aa}  

\usepackage{graphicx}
\usepackage{txfonts}
\usepackage{gensymb}
\usepackage{mathtools}
\usepackage{amsmath}
\usepackage{mathrsfs}
\usepackage{natbib}
\usepackage{booktabs}
\usepackage{array}
\usepackage{multirow}
\usepackage{makecell} 
\usepackage{tabularx}
\usepackage{hyperref}
\usepackage{siunitx}
\usepackage{xspace}
\newcommand{\MdotIBH}{\,$\dot{M}_{\rm IBH}$\xspace}

\newcommand{\MdotEdd}{\,$\dot{M}_{\rm Edd}$\xspace}

\newcommand{\kms}{~km\,s$^{-1}$}   
\newcommand{\ergs}{~erg\,s$^{-1}$}

\begin{document}

   \title{Probing the detectability of electromagnetic signatures from Galactic isolated black holes }

   \author{J. R. Martinez
          \inst{1,2}
          \and
          V. Bosch-Ramon\inst{3}
          \and
          F. L. Vieyro\inst{2,4}
          \and
          S. del Palacio\inst{5}
          }

   \institute{Facultad de Ciencias Exactas, UNLP,
              Calle 47 y 115, CP(1900), La Plata, Buenos Aires, Argentina.\\
            \email{jmartinez@iar.unlp.edu.ar}
         \and
             Instituto Argentino de Radioastronom\'ia (CCT La Plata, CONICET), C.C.5, (1894) Villa Elisa, Buenos Aires, Argentina.
        \and
            Departament de F\'isica Qu\'antica i Astrof\'isica, Institut de Ci\'encies del Cosmos (ICCUB), Universitat de Barcelona (IEEC-UB), Mart\'i i Franqu\`es 1, E-08028 Barcelona, Spain.
        \and
            Facultad de Ciencias Astronómicas y Geofísicas, Universidad Nacional de La Plata, Paseo del Bosque s/n, 1900 La Plata, Argentina           
        \and
             Department of Space, Earth and Environment, Chalmers University of Technology, SE-412 96 Gothenburg, Sweden.
             }
   \date{}

% \abstract{}{}{}{}{} 
% 5 {} token are mandatory
 
  \abstract
   {A large number of isolated stellar-mass black holes (IBHs) are expected to populate the Galaxy. However, only one has been confirmed by the analysis of a microlensing event, and no confirmed emission detection from an IBH has been reported so far.}
   {We analysed the detectability of electromagnetic signatures from IBHs moving in the Galaxy. 
   }
   {We considered accretion from the interstellar medium onto an IBH and assumed the formation of an outflow. We then semi-analytically modelled the accretion process and the interaction of the outflow with the surrounding medium on large scales, including mechanical feedback on the accretion process. Furthermore, we also (semi-)analytically calculated the emission from three different regions: the accretion region, the thermal and the non-thermal radiation from the outflow-medium interaction structure, and the non-thermal emission of relativistic particles that diffuse in the surrounding medium.}
   {Our results show that multi-wavelength emission associated with Galactic IBHs can be detected in systems moving through
   a very dense medium such as the core of a molecular cloud. In particular, thermal emission from accretion could be observed in the mid-infrared and in hard X-rays with current and forthcoming observatories. Thermal and non-thermal emission from the outflow-medium shock could also be detected in the radio and millimetre ranges. Moreover, detection of the emission from particles diffusing in a dense medium could be feasible in $\gamma$-rays. Applying our model to the IBH associated with the gravitational microlensing event MOA-2011-BLG-191/OGLE-2011-BLG-0462, we inferred that radio and infrared detection of the IBH is plausible. Also, we derived that IBHs could be modest Galactic cosmic ray contributors, potentially reaching a $\sim 1$\% contribution at $E\gtrsim 1$~PeV. Finally, by extending our model to primordial black holes, we conclude that efficient leptonic acceleration in their outflow–medium interactions would rule them out as a major dark matter component.}
  {}


   \maketitle
%
%-------------------------------------------------------------------
\section{Introduction}\label{Sec:intro}

Isolated black holes (IBHs) are stellar-mass black holes that lack a companion. The existence of these objects is expected for several reasons. On the one hand, about 30\% of massive stars that end their lives as black holes are isolated \citep{Sana_2012}. On the other hand, IBHs can form by the merger of binary systems \citep{Zhang_2001, Tutukov_2011}. In addition, supernova explosions can eject a component from a binary at high speeds, breaking up the system and creating an isolated object that moves away from its birthplace at velocities of tens to hundreds of kilometres per second \cite[e.g.][and references therein]{Sahu_2022}. Based on the mass evolution of the Galaxy and the mass distribution of stars, it is estimated that there are of the order of $10^8$ IBHs in the Milky Way \citep{van_den_Heuvel_1992}. Although the distribution of these objects is uncertain, this number implies an estimated mean density of IBHs of $5 \times 10^5$~kpc$^{-3}$ \citep{Fender_2013}. We note that multiple black hole systems may not differ too strongly from the IBH scenario, and orbital motion may even enhance the effects studied here, but the analysis in that case is more complicated and thus left for future work.

Despite the large number of hypothetical systems, the first detection of an IBH occurred recently. \cite{Sahu_2022} reported the detection of an IBH through the analysis of the microlensing event MOA-2011-BLG-191/OGLE-2011-BLG-0462. From the size of the Einstein ring and the characteristics of the light curve, the authors inferred a black hole mass of $M_{\rm IBH} = (7.1\pm 1.3)$~M$_\odot$, a peculiar velocity of $v_{\rm IBH} \approx 45$\kms (although  in a more recent study \citealt{Sahu_2025} derived a value of $v_{\rm IBH} \approx 51$\kms), and a distance of $(1.58\pm 0.18)$~kpc. This detection, in conjunction with the advent of new observational catalogues such as the GAIA Data Release 3 \citep{Gaia3}, bodes well for the future of IBH research. \cite{Mereghetti_2022} did not detect X-ray emission from this object despite different models predicting detectable broadband radiation from IBHs \citep[e.g.][]{Agol_2002,Barkov_2012,Fender_2013,Abaroa_2024,Kin_2025}. Thus, the detection of electromagnetic emission from IBHs will likely represent a significant challenge in the coming years.

In this work, we addressed the detectability of electromagnetic signatures of IBHs in the Milky Way. For this purpose, we employed semi-analytical dynamical and radiative models to predict the broadband thermal and non-thermal emission from IBHs, and compared the predictions with the sensitivity of current and forthcoming instruments. The paper is organised as follows: In Sect.~\ref{Sec:physics} we describe the physical properties of the scenario studied. In Sect.~\ref{Sec:Model} we present the model we developed. In Sect.~\ref{Sec:Results} we present the main results of our work, which includes the application of our model to the system  MOA-2011-BLG-191/OGLE-2011-BLG-0462 and a brief discussion of our results in the context of primordial black holes (PBHs). Finally, in Sect.~\ref{Sec:Conclusions} we summarise our main conclusions.
%-------------------------------------------------------------------
%-------------------------------------------------------------------

\section{Physical scenario}\label{Sec:physics}

An IBH moving supersonically with respect to its surrounding medium accretes under a cylindrical accretion regime, although medium inhomogeneities and anisotropies and outflows can lead to the breaking of this symmetry. Material falling within the IBH radius of influence forms an accretion column behind it. The size of this cross-section depends on the mass and velocity of the IBH, as well as the sound speed of the surrounding medium \citep{Bondi_1952, Fujita_1998}:
\begin{equation}
    r_{\rm acc} \approx \frac{2GM_{\rm IBH}}{v_{\rm IBH}^2 + c_{\rm s}^2}\,,
    \label{Eq:r_acc}
\end{equation}
with $G$ being the gravitational constant. Moreover, the accretion rate also depends on the medium density, $\rho_{\rm med}$, as \citep{B-H_1944}:
\begin{equation}
    \dot{M}_{\rm IBH} \sim \lambda_{\rm acc}\,\pi\,r_{\rm acc}^2\,\rho_{\rm med}\,v_{\rm IBH} \sim \lambda_{\rm acc} \frac{4\pi G^2 M_{\rm IBH}^2\,\rho_{\rm med}}{\left(v_{\rm IBH}^2+c_{\rm s}^2\right)^{3/2}}\,,
    \label{Eq:Mdot_IBH}
\end{equation}
with $\lambda_{\rm acc} < 1$. This parameter takes into account all the physical processes that can reduce the accretion towards the IBH. These processes can operate on scales of the order of the accretion radius, as in the case of viscosity effects or radiation feedback \citep[see, e.g.][]{P&R_2013,Scarcella_2021}, as well as on larger scales. The presence of outflows can also reduce accretion by effectively increasing the pressure of the accreted medium by mechanically interacting with it \citep{B-R_2020,B-R_2021}. This phenomenon is known as mechanical feedback and has been explored in other contexts \citep[e.g.][]{Li_2020,Chen_2023}. Since in this work we focus on the interaction via mechanical feedback and treat the effects near the black hole phenomenologically, we set $\lambda_{\rm acc}$ = 0.1, a plausible value according to \cite{B-R_2021}.

Isolated black holes are anticipated to possess low-to-modest mass accretion rates, which give rise to the formation of an advection-dominated accretion flow \citep[ADAF; e.g.][]{Fujita_1998}. This accretion regime allows for the ejection of outflows in the form of winds or collimated jets \citep[see, e.g.][and references therein]{ Blandford_1999,Yuan_2014,Gutierrez_2021}. Moreover, even if the infalling material does not have enough angular momentum to develop an ADAF, powerful jets can still be launched \citep{Barkov_2012_b}. We assumed that a forward outflow is launched forming an angle $\theta$ with the direction of motion of the IBH, while a rear outflow is launched towards the opposite direction (see a scheme in Fig.~\ref{fig:IBH}). The trajectories of the outflows are almost ballistic until they are deflected by the medium ram pressure \citep{B-R_2020}. This interaction depends on $\theta$. For an angle $\theta \lesssim 30\degree$, the collision can be considered head-on because the shocked outflow sound speed is higher than its speed parallel to the shock, and the structure is similar to a bow shock. For a collision angle of $30\degree \lesssim \theta \lesssim 90\degree$, the impact is oblique enough to be qualitatively more similar to a quasi-lateral interaction, and only the forward outflow momentum component in the direction of motion of the IBH is halted; the rear outflow is also deflected due to the lateral impact of the medium, but more obliquely. Finally, for $ 90\degree-\chi \lesssim \theta \le 90\degree$, with $\chi$ being the half-opening angle of the outflow, both outflows can be taken as perpendicular to the IBH motion and are shocked laterally. In both the oblique and the perpendicular scenarios, the material from both shocked outflows convect away through a flow tube. It is worth mentioning that oblique outflow production likely requires perturbations that break the axial symmetry of the accretion process. This can be attributed to several phenomena, as discussed in \cite{B-R_2020}. On the one hand, the movement of the surrounding material results in the relative velocity between the IBH and the medium being non-uniform. Secondly, the symmetry can also be broken by frame-dragging in a rotating IBH scenario. Moreover, the compression of a local or large-scale ISM magnetic field can generate an inhomogeneous pressure that deflects the launched material from the symmetry axis.

The radiation from the accreting flow may be significant. In addition, the outflows can produce or inject non-thermal particles into different regions through shock and escape processes, respectively. These particles may, in turn, generate detectable amounts of multi-wavelength radiation. On the other hand, thermal emission from the outflow-medium interaction can also be important. In the next section we introduce an analytical model to characterise all this emission.

%--------------------------------------------------------------------
%====================================================================

%--------------------------------------------------------------------
%====================================================================
%--------------------------------------------------------------------

\section{Model}
\label{Sec:Model}

\begin{figure}[t]
    \centering
    \includegraphics[width=0.49\textwidth]{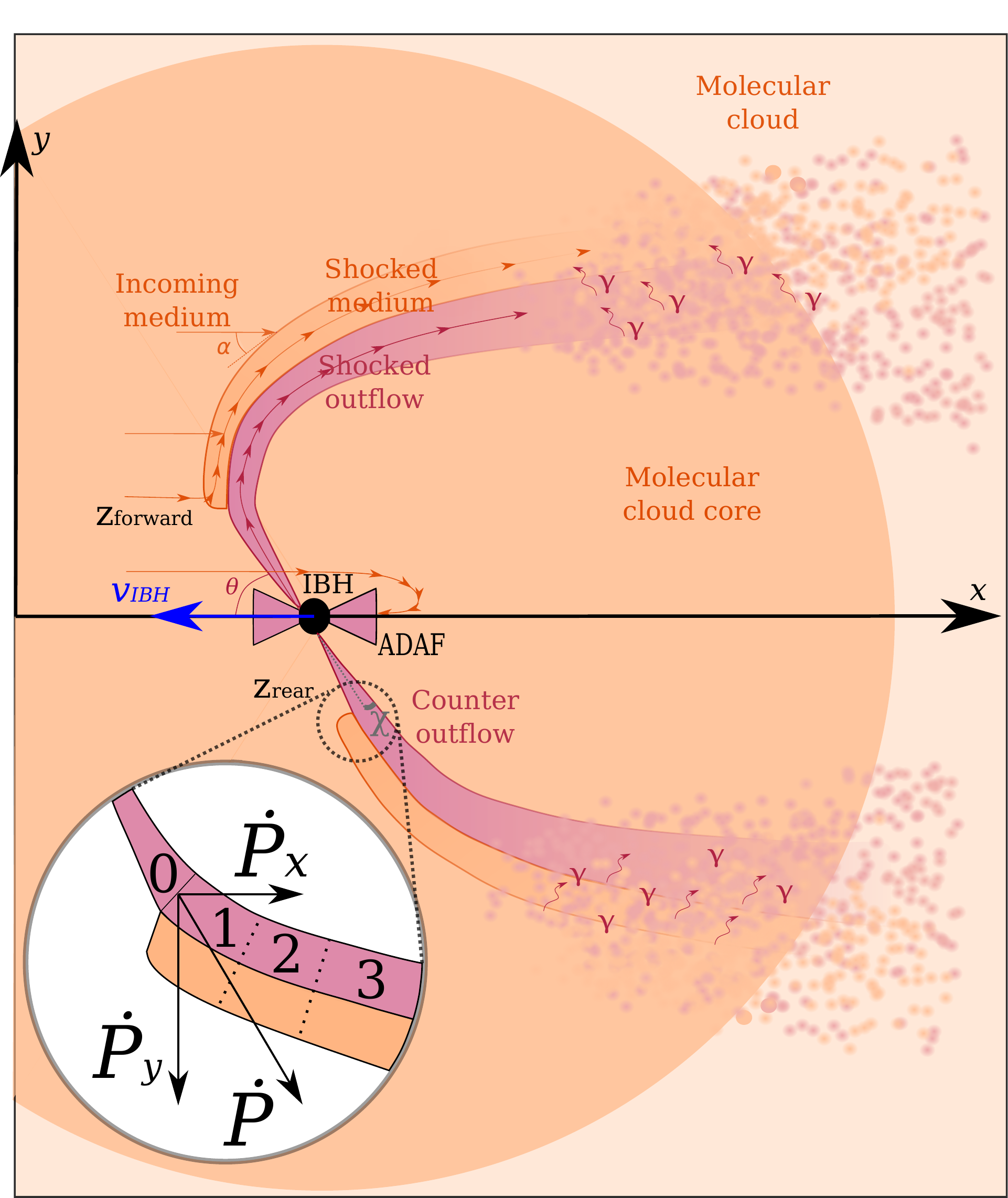}
    \caption{Sketch (not to scale) of the modelled system. The IBH accretes and an ADAF forms. An outflow is launched forming an angle $\theta$ with the IBH direction of motion. The forward outflow is deflected by the medium pressure at $z_{\rm forward}$, while the rear outflow is deflected at $z_{\rm rear}$. Particles are accelerated via diffusion mechanisms at the deflection points. The shocked material convects away until being disrupted, and energetic particles are injected into the molecular cloud core. Finally, these particles diffuse through the core and the rest of the molecular cloud. We also zoom in the region close to the deflection point of the rear outflow. We show the momentum rate vector of the outflow and we schematise the first cells of the interaction structure with the labels 1, 2 and 3. The unshocked outflow is labelled with a 0.}
    \label{fig:IBH}
\end{figure}

Isolated black holes are expected to exhibit a low accretion rate. In particular, IBHs within mediums with densities similar to the average density in the Galactic disc ($\sim$0.1--1~cm$^{-3}$; \citealt{fer01}) should emit negligible radiation. This prediction is consistent with the present lack of detections of IBH electromagnetic signatures. Therefore, we focused on a potential scenario in which an IBH crosses a molecular cloud core, as illustrated in Fig~\ref{fig:IBH}. These regions are characterised by high densities of $n_{\rm med} \sim 10^5$~cm$^{-3}$, low temperatures of $T_{\rm med} = 10$~K, and sizes of approximately $0.1$~pc \citep[e.g.][]{fer01,Stahler_Palla_2004}. We also considered an IBH with a mass of $M_{\rm IBH} = 10$~M$_\odot$, which could be a typical value for stellar-mass black holes \citep{Agol_2002}. In addition, we set the velocity of the IBH at $v_{\rm IBH} = 30$~km s$^{-1}$, similar to the average value for massive stars \citep{Agol_2002, Fender_2013}. Plugging these values into Eq.~\ref{Eq:Mdot_IBH}, we obtained an accretion rate of \MdotIBH $\approx 2 \times 10^{16}$~g\,s$^{-1} \sim 10^{-3}$\MdotEdd, where \MdotEdd $\approx 1.4 \times 10^{19}$~g\,s$^{-1}$ is the Eddington accretion rate (related to the Eddington luminosity $L_{\rm Edd}=1.3 \times 10^{38} \left(M_{\rm IBH}/{\rm M}_\odot\right)$~erg s$^{-1}$ as $\dot M_{\rm Edd}=10\,L_{\rm Edd}/c^2$). 

With regard to the outflow, we set $\theta=60\degree$, the mean angle for a random distribution of outflow orientations, which corresponds to an oblique interaction. Our results exhibit only minor variations when we alter the $\theta$-value in the oblique interaction regime. The collimation angle plays a smaller role in mechanical feedback \citep{B-R_2021}. For the sake of simplicity, we modelled collimated outflows with an intermediate half-opening angle of $\chi = 0.2$. Smaller values of $\chi$ should not strongly change the results, while significantly larger values would further reduce accretion due to mechanical feedback, as they are more akin to the head-on interaction regime. We summarise the selected parameters in Table~\ref{table:IBH_parameters}.

\begin{table}[t]    
    \centering
    \caption[]{Model parameters.}
    \begin{tabular}{l l c}
    \hline\hline\noalign{\smallskip}
    \multirow{3}{*}{Parameter}     & \multirow{3}{*}{Symbol}     & \multirow{3}{*}{Value} \\ \\ \hline \\
    IBH velocity& $v_{\rm IBH}$ & 30 km s$^{-1}$\\
    IBH mass& $M_{\rm IBH}$ & 10 M$_\sun$\\
    Accretion parameter & $\lambda_{\rm acc}$ & 0.1 \\
    Unshocked outflow velocity & $v_{\rm out}$ & $0.5\,c$ \\
    Outflow inclination & $\theta$ & 60\degree \\
    Outflow half-opening angle & $\chi$ & 0.2 \\
    \hline\noalign{\smallskip}
    Magnetic field parameter & $\eta_{\rm B}$ & 0.1 \\
    Non-thermal luminosity parameter & $\eta_{\rm NT}$ & 0.1 \\
    \hline\noalign{\smallskip}
    Distance & $d$ & 2~kpc \\
    \hline\noalign{\smallskip}
    Medium density & $n_{\rm med}$ & 10$^5$~cm$^{-3}$ \\
    Medium temperature & $T_{\rm med}$ & 10~K \\
    
    \bottomrule
    \end{tabular}
    \tablefoot{Mass and velocity of the IBH are taken as typical values for stellar-mass black holes according to \cite{Agol_2002} and \cite{Fender_2013}. The unshocked outflow velocity is derived from the relations given in \cite{HeinzGrimm2005}. The outflow inclination and half-opening angles are taken from \cite{B-R_2021}. The quantities $\lambda_{\rm acc}$, $\eta_{\rm B}$ and $\eta_{\rm NT}$ are free parameters of the model (see discussion in Sects.~\ref{subsec:res_out} \& \ref{Subsec:lambda_acc}). Medium properties are taken from \cite{fer01} and \cite{Stahler_Palla_2004}.}
\label{table:IBH_parameters}
\end{table}

%---------------------------------------------------------------------------------------------------------------------------------------------------------------------------------
\subsection{Ambient medium}\label{Subsec:molecular_cloud}

Molecular clouds consist of gas and dust heated by stellar radiation to temperatures of tens of Kelvin. Due to their high density and the presence of dust, these regions produce significant absorption of radiation in the near-IR, optical, ultraviolet and X-rays. Moreover, the dust is a bright emitter in the far-IR. To calculate the dust emission, we modelled it as a modified blackbody emitter of intensity:
\begin{equation}
    I_\nu\left(\nu\right)= \tau_0 \left(\frac{\nu}{\nu_0}\right)^\beta B_\nu(\nu,T),
    \label{Eq:I_dust}
\end{equation}
where $\tau_0$ is the dust optical depth at a reference frequency $\nu_0$, $\beta$ is the opacity spectral index, and $B_\nu$ is the Planck function. According to \cite{Planck_2014}, $\beta$ takes different values between millimetre wavelengths and the far infrared (IR): $\beta = 1.54$ for $\nu < 353$~GHz, and $\beta = 1.8$ for $\nu > 353$~GHz. Furthermore, \cite{Planck_2014} estimated a dust optical depth of $\tau_0 \approx 4 \times 10^{-4}$ at $\nu_0 = 353$~GHz within dense cores.

In addition to considering the emission from the core, it is essential to take into account the radiation absorption and extinction processes generated by this dense region. Dust grains cause extinction of radiation at wavelengths comparable to or smaller than their characteristic sizes. They absorb visible and ultraviolet radiation, with an opacity peak at $\lambda \approx$~\SI{2200}{\angstrom} due to the presence of silicates. Absorption at the IR is not trivial to account for because of the possible development of ice mantles \citep{Whittet_1988}. Nevertheless, the extinction decreases with wavelength as $A_\lambda \propto \lambda^{-1.8}$ in this band, and molecular clouds become transparent to the continuum in the mid-IR, above $\lambda \approx1$~$\mu$m \citep{Lada_1999}. Observing sources inside the core from the near-IR to the ultraviolet represents a challenging task. In particular, this hinders the detection of emission lines such as H$_\alpha$, which can be significant in interstellar shocks with velocities of tens of \kms. Additionally, the cloud absorbs photons with energies above the hydrogen ionisation energy and up to the soft X-rays. Using the formalism detailed by \cite{Morrison_McCammon_1983}, we obtained that the spectra  associated with the IBHs are absorbed up to energies of approximately $\sim 1$~keV. Regarding absorption outside the cloud core, this can be neglected at a distance of $\sim 2$~kpc.

Considering the absorption and emission processes involved, we focused our study on specific regions of the electromagnetic spectrum. We inferred that the most promising bands for the detection of IBHs within dense regions are the radio-to-millimetre band, the mid-IR, and at high energies, from hard X-rays to $\gamma$-rays.

%---------------------------------------------------------------------------------------------------------------------------------------------------------------------------------
\subsection{Accretion emission}\label{Subsec:ADAF_disk}

An IBH accretion flow with a mass rate of approximately $10^{-3}$\MdotEdd and sufficient angular momentum develops an ADAF. The low density characteristic of these structures results in inefficient interactions between particles and small radiation losses, and the flow dynamics is dominated by advection towards the compact object. Moreover, electrons and protons do not thermalise at the same temperature \citep{Narayan_1995}. The cooling mechanisms of electrons are more efficient than those of protons, while the viscosity dissipation process heats protons more efficiently. Consequently, protons reach higher temperatures than electrons.

The high proton temperature in the ADAF gives rise to a high pressure despite the modest densities, resulting in the formation of a geometrically thick, optically thin accretion structure with a spectrum that is not that of a multi-color black body.  The primary radiation processes are synchrotron emission and bremsstrahlung, both modified by Comptonisation. Synchrotron photons are upscattered via Comptonisation to energies as high as 100 keV to 1 MeV. At high accretion rates, this component can dominate the spectrum, whereas at very low values of the accretion rate, Comptonisation remains weak \citep{Yuan_2014}. To calculate this spectrum, we applied the model developed by \cite{Gutierrez_2021} considering electron emission. We have adopted standard values of the ADAF parameters, as the one adopted in \cite{Gutierrez_2021} as reference values; a detailed study of the dependence of the emission on the ADAF parameters can be found in the literature \citep{Narayan_1995,Gutierrez_2021}.

%---------------------------------------------------------------------------------------------------------------------------------------------------------------------------------
\subsection{Outflow-medium interaction}\label{Subsec:outflow}

\subsubsection{Outflow properties}
\label{Subsec:v_out}

The low-accretion scenario has been studied more extensively in binaries, but the physics near the compact object should also resemble that of an IBH. According to \cite{HeinzGrimm2005}, a black hole in a binary system undergoes accretion under the ADAF regime when the accretion rate is below a transition value of $\dot{M}_{\rm tran} \sim 0.01$\MdotEdd. Defining the transition luminosity as $L_{\rm X, tran} \sim 0.1 \dot{M}_{\rm tran}c^2$, the X-ray luminosity of the ADAF can be expressed as:
\begin{equation}
  L_{\rm X,ADAF} \sim L_{\rm X,tran}\,\left(\frac{\dot{M}_{\rm IBH}}{\dot{M}_{\rm tran}} \right)^2.
  \label{Eq:L_X_ADAF}
\end{equation}
Furthermore, by employing Eq.~(4) from \cite{HeinzGrimm2005} we derived that an IBH with an accretion rate of \MdotIBH$\approx 2 \times 10^{16}$~g\,s$^{-1}$ can launch collimated outflows with a total power of $L_{\rm out,tot} \sim 2 \times 10^{36}$\ergs. Given this estimate, we assumed a total luminosity of $L_{\rm out} \approx 10^{36}$\ergs.

The initial velocity ($v_{\rm out}$) and mass rate of each outflow before being shocked and affected by the medium lateral impact are related to the power of each component as:
\begin{equation}
    L_{\rm out,f} = L_{\rm out,r} = 0.5 L_{\rm out}\approx 0.5\dot{M}_{\rm out}\,\left(\gamma_{\rm out}-1\right)c^2\,,
    \label{Eq:L_out}
\end{equation}
where the subscript f(r) stands for forward(rear), and with $\gamma_{\rm out}$ being the outflow Lorentz factor. These quantities are needed to derive the outflow momentum rate and characterise the medium-outflow interaction. The initial outflow mass rate is fixed by $\epsilon = \left(\dot{M}_{\rm out}/\dot{M}_{\rm IBH} \right)$. A possible value for this quantity may be $\epsilon\sim 0.5$, which would mean $v_{\rm out}\approx 0.5\,c$ for the power obtained above. We note that a much faster outflow may in principle be possible, reducing its mass rate, as the actual constraints are only the power and $\epsilon<1$ in the unshocked region of the outflows. We adopted, however, a moderately relativistic $v_{\rm out}$ consistent with what is expected for persistent jets from stellar-mass black holes \citep{Saikia_2019}. 

%--------------------------------------------------------------------
%--------------------------------------------------------------------

\subsubsection{Shocked outflow thermodynamics}\label{subsubsec:Thermo_RS}

We considered that the outflows propagate and expand freely until they are confined by the ambient ram pressure. At that point, outflow deflection becomes important, in a scenario similar to that proposed by \cite{B-R_2016} and \cite{Barkov_2022} for the interaction between jets and stellar winds in high-mass microquasars. After the deflection point, the shocked outflows are assumed to be separated from the shocked incoming medium by contact discontinuity surfaces. In the forward outflow, the medium ram pressure roughly becomes in balance with the component of the outflow momentum rate along the IBH motion, whereas in the rear component, it is the outflow lateral ram pressure that becomes in balance with the thermal pressure of the shocked medium, similar in the deflection region to the unshocked medium ram pressure. To calculate the thermodynamics of the shocked gas and the resulting emission, we developed a multi-zone model based on \cite{del_Palacio_2018} and \cite{Martinez_2022}. Following the approach of previous works on collimated outflows, we modelled the shocked medium and outflows as 1-D structures \citep{Molina_2018,Molina_2019}. Each structure is discretised into multiple cells, and at each cell we assigned thermodynamical and geometrical properties (size and lateral and cross-sectional areas). The relevant component of the forward outflow ram pressure and the rear outflow lateral ram pressure before the deflection are:
\begin{equation}
    P_{\rm out} =
    \begin{dcases}
        \frac{L_{\rm out}\,\gamma_{\rm out}\,\beta_{\rm out}\cos{\left(\theta\right)^2}}{c\left(\gamma_{\rm out}-1\right)\,\pi \tan^2{\left(\chi\right)}\,z^2} & {\rm forward\,outflow}\\
        \frac{L_{\rm out}\,\gamma_{\rm out}\,\beta_{\rm out}}{c\left(\gamma_{\rm out}-1\right)\,\pi\,z^2} & {\rm rear\,outflow}
        ,
    \end{dcases}
    \label{Eq:P_out}
\end{equation}
with $z$ being the coordinate along the outflow from the launching position and $\beta_{\rm out} = v_{\rm out}/c$. On the other hand, the ambient ram pressure is $P_{\rm med} = \rho_{\rm med}v_{\rm IBH}^2$. Then, the outflows deflection points, at which $P_{\rm out} = P_{\rm med}$, are located at:
\begin{equation}
    z_{\rm out} \approx
    \begin{dcases}
     \sqrt{\frac{L_{\rm out}\,\gamma_{\rm out}\,\beta_{\rm out}\cos^2{\left(\theta\right)}}{c\left(\gamma_{\rm out}-1\right)\,\pi \tan^2{\left(\chi\right)}\,\rho_{\rm med}\,v_{\rm IBH}^2}} & {\rm forward\,outflow} \\ \\
    \sqrt{\frac{L_{\rm out}\,\gamma_{\rm out}\,\beta_{\rm out}}{c\left(\gamma_{\rm out}-1\right)\,\pi\,\rho_{\rm med}\,v_{\rm IBH}^2}} & {\rm rear\,outflow}\,.
    \end{dcases}
    \label{Eq:z_0}
\end{equation}

Right before the deflection point, the unshocked density of each component is:
\begin{equation}
    \rho_{\rm out} \approx
    \frac{0.5L_{\rm out}}{\gamma_{\rm out}\left(\gamma_{\rm out}-1\right)v_{\rm out}c^2 A_{\rm out}}\,,
    \label{Eq:rho_out_0}
\end{equation}
with $A_{\rm out} = \pi\left(z_{\rm out}\tan{\left(\chi\right)}\right)^2$ being the outflow cross-section. Immediately after the shock, Rankine-Hugoniot jump conditions determine the flow dynamics. However, some outflow-shocked medium mixing is likely to take place within a short region after the shock, leading to mass loading in the shocked outflow tube. For this reason, we do not employ Rankine-Hugoniot conditions to compute the thermodynamic properties at the first cell. To obtain the post shock density\footnote{We use the subscript 1 for quantities at the first cell.}, $\rho_1$, we consider that the mass-loading process implies conservation of energy flux across the shock, but neither momentum nor mass are conserved, and velocity $v_1$ is taken as a free parameter. The outflow-medium mixing also means that a non-relativistic flow model is a good approximation beyond the deflection region. From energy conservation,
\begin{equation}
    \left(0.5\rho_1 v_1^3 + \frac{\gamma_{\rm ad}}{\gamma_{\rm ad}-1}P_1 v_1 \right)A_{\rm out} = 0.5L_{\rm out},
    \label{Eq:cons_E}
\end{equation}
we obtained:
\begin{equation}
    \rho_1 = \frac{2}{v_1^3}\left(\frac{0.5L_{\rm out}}{A_{\rm out}} - \frac{\gamma_{\rm ad}}{\gamma_{\rm ad}-1} 
P_1v_1\right),
    \label{Eq:rho_1}
\end{equation}
with $\gamma_{\rm ad}$ being the adiabatic index of the gas. Taking into account that $P_1 = P_{\rm out}$, Eq.~(\ref{Eq:rho_1}) allowed us to calculate the densities right after the shock assuming a post-shock velocity. Mixing after the deflection points leads to a higher density, and thus to a lower velocity. Our phenomenological approach consisted in assuming a value for $v_1$, and deriving $\rho_1$ from it. In particular, we assumed $v_1 = 0.1c$, resulting in a high overdensity ($\rho_1 \sim 100\,\rho_{\rm out}$) for both components of the outflow. We point out that the value assumed for $v_1$ allowed us to neglect emission relativistic effects such as Doppler boosting or beaming, and the resulting emission is thus not severely affected by variations of $v_1$ of up to a few times its assumed value. For completeness, we also addressed the scenario without mixing, in which Rankine-Hugoniot jump conditions determine the thermodynamics at the first cell, and showed that the impact on the results are moderate (see Sect.~\ref{subsec:res_out}). We note that the mixing process is not instantaneous, and that the real scenario should lie between the mixing and non-mixing cases.

Once the outflow plasma is shocked, its thermal pressure becomes roughly equal to the outflow ram pressure balanced by that of the medium, which means that the post-shock pressure is $P_1 = P_{\rm med}$. The pressure in the subsequent positions is determined by the impact of the medium further downstream. If the shocked medium is subsonic, this pressure at the $l$-th cell is $P(l) = \rho_{\rm med}\,v_{\rm IBH}^2\,\sin{\left(\alpha(l)\right)}$, while it is $P(l) = \rho_{\rm med}\,v_{\rm IBH}^2\,\sin^2{\left(\alpha(l)\right)}$ when the shocked medium becomes supersonic, with $\alpha$ being the angle between the direction of motion of incoming medium (in the IBH rest frame) and the shocked outflow surface (see Fig.~\ref{fig:IBH}). Knowing the pressure cell by cell, and assuming that the shocked outflow behaves as an ideal gas with adiabatic index $\gamma_{\rm ad} = 5/3$, one can calculate the mass density in the cell:
\begin{equation}
    \rho(l) = \rho_1\left(\frac{P(l)}{P_1}\right)^{1/\gamma_{\rm ad}}.
    \label{Eq:rho(l)}
\end{equation}

The shocked outflow accelerates as it convects away from the deflection point; no more mixing occurs until it is fully disrupted. Employing Bernoulli's equation, we calculated the velocity at each position:
\begin{equation}
    v(l) = \sqrt{v_1^2 + 2 \frac{\gamma_{\rm ad}}{\gamma_{\rm ad}-1} \left(\frac{P_1}{\rho_1}-\frac{P(l)}{\rho(l)}\right)}
    \label{Eq:v_shocked_out}.
\end{equation}
This allowed us to compute the cross-section of the tube at each position by applying mass conservation beyond the first cell, where the mass-loading process has already occurred:
\begin{equation}
    A(l) = A_1\,\frac{\rho_1\,v_1}{\rho(l)\,v(l)}\,.
\end{equation}

Finally, with regard to the magnetic field, we assumed that the magnetic pressure at the deflection point is a fraction $\eta_{\rm B}$ of the thermal\footnote{We note that the thermal pressure includes the pressure contribution due to non-thermal particles, and the dynamical role of the magnetic pressure is neglected.} pressure:
\begin{equation}
    \frac{B_1^2}{8\pi} = \eta_{\rm B}P_1,
    \label{Eq:B_1}
\end{equation}
and considered a reference value of $\eta_{\rm B} = 0.1$ \citep[see, e.g.][for a discussion about the value of this parameter in the context of stellar bow shocks]{del_Palacio_2018,Benaglia_2021}. The presence of a strong magnetic field could affect fluid compression. However, for the adopted value of $\eta_{\rm B}$, the magnetic pressure is low, and we therefore neglect its effects for simplicity. Finally, to calculate the magnetic field in the subsequent cells, we assumed for simplicity a perpendicular magnetic field and magnetic flux conservation, yielding the following:
\begin{equation}
    B(l) = B_1 \sqrt{\frac{\rho(l)\,v_1}{\rho_1\,v(l)}}.
\end{equation}

%====================================================================================================
\subsubsection{Shape of the interaction structure}
\label{Subsec:CD}

To calculate the shape of the entire shocked flow structure in the plane containing $\vec{v_{\rm IBH}}$ and $\vec{v_{\rm out}}$, we effectively treated the structure as uniform in the direction perpendicular to that plane. Furthermore, we neglected the effect of the shocked medium circumventing the outflow tube, which was also assumed to be roughly planar on its side facing the incoming medium. We then discretised the tube in multiple cells of length $dz(l)$, with the position of the $l-$th cell in the $xy$ plane being:
\begin{align}
    x(l) &= x(l-1) + dx(l) = x(l-1) + \frac{\dot{P}_x(l)\,dz(l)}{\dot{P}(l)} \\
    y(l) &= y(l-1) + dy(l) = y(l-1) + \frac{\dot{P}_y(l)\,dz(l)}{\dot{P}(l)},
    \label{Eq:xy_coordinates}
\end{align}
where $\vec {\dot{P}}(l)$ is the total momentum rate of the shocked structure (i.e. including both shocked medium, and shocked outflow -tube-), determined in each position by the ram pressure imparted by the incoming medium. For the two relevant components of the vector $\vec{\dot{P}}(l)$:
\begin{align}
    \dot{P}_x(l) &= \dot{P}_x(l-1) + d\dot{P}_x(l) = \dot{P}_x(l-1) + P(l)\,ds(l) \\
     \dot{P}_y(l) &= \dot{P}_y(l-1) + d\dot{P}_y(l) = \dot{P}_y(l-1),
     \label{Eq:dotP}
\end{align}
with $ds(l) = 2\,dz(l)\,h(l)\sin{\left(\alpha(l)\right)}$ being the projection of the structure surface in the direction of ${\bf v_{\rm IBH}}$, and $h(l)$ the radius of the flow tube. In addition, the initial conditions at the first cell are:
\begin{align}
    \dot{P}_x(1) &= A_{\rm out}\,\left(\rho_1 v_1^2 + P_1\right)\,\cos{\left(\theta\right)} \\
    \dot{P}_y(1) &= A_{\rm out}\,\left(\rho_1 v_1^2 + P_1\right)\,\sin{\left(\theta\right)}.
    \label{Eq:dotP_initial}
\end{align}

We remark that if the shocked material behaved as a laminar flow along the tube made of shocked outflow and surrounded by shocked, stable medium, the tube would last until its pressure equalises with the medium thermal pressure. Nonetheless, a variety of effects predict premature disruption of the outflow. On the one hand, a dense layer of unstable plasma can be created behind the radiative forward shock, similar to the radiative phase scenario of a supernova remnant \citep{Cioffi_1988,Blondin_1998,Bandiera_2004}. On the other hand, the shocked outflow convects at a velocity $\sim v_1$, while the shocked medium convects at a velocity $\sim v_{\rm IBH}$. This large velocity jump triggers Kelvin-Helmholtz instabilities. The combined effect of these instabilities leads to the eventual disruption of the tube not far from the deflection point, as supported by numerical simulations \citep{B-R_2021,Barkov_2022}. We therefore tracked the outflow material until it travels a distance $\sim 15\,z_{\rm out} \approx 0.05$~pc, point at which it was assumed to disrupt and mix with the surrounding medium. Although this disruption could take place before, the main results presented in this paper would remain valid. Both the acceleration of non-thermal particles and the majority of the emission from the interaction structure are concentrated in a region close to the deflection points. Additionally, non-thermal particles could also be accelerated in the disruption region (i.e. the region at $z \sim 15\,z_{\rm out}$, where the outflow and the medium are completely mixed\footnote{This region should not be confused with the one where the outflows are deflected by the shocked medium at the onset of the tube, which may also result in the process of mixing.}). However, we do not consider this effect, as such detail in the acceleration process is beyond the scope of this work.

%--------------------------------------------------------------------
%-------------------------------------------------------------------

%--------------------------------------------------------------------
\subsubsection{Non-thermal particles}\label{Subsubsec:NT_particles}

Non-thermal electrons and protons could be accelerated in both the forward outflow and the rear outflow deflection regions via diffusive shock acceleration. Acceleration further downstream may take place, but for simplicity we assumed an accelerator located at the deflection points. We considered that the power injected into these particles is a fraction $\eta_{\rm NT}$ of the kinetic power injected perpendicularly into the shocks:
\begin{equation}
    L_{\rm inj, NT} =
    \begin{dcases}
        \eta_{\rm NT}\,0.5\,L_{\rm out}\cos^2{\left(\theta\right)} & {\rm forward\,outflow} \\
        \eta_{\rm NT}\,0.5\,L_{\rm out}\tan^2{\left(\chi\right)} & {\rm rear\,outflow}.
    \end{dcases}
    \label{Eq:L_inj,NT}
\end{equation}
In turn, we assumed that 90\% of this power is injected into protons, while the remaining 10\% is injected into electrons. The quantity $\eta_{\rm NT}$ and the above percentages are free parameters, and the corresponding radiated luminosities scale linearly with them; different configurations will favour either leptonic or hadronic emission depending on these percentages. For the injection function of non-thermal particles, we assumed a power law with an exponential cutoff:
\begin{equation}
    Q(E) \propto E^{-2}\,e^{\left(-E/E_{\rm max}\right)},
    \label{Eq:Q_inj}
\end{equation}
 where $\left[Q\left(E\right) \right]=$~erg$^{-1}$\,s$^{-1}$, typical for  diffusive shock acceleration \citep{Drury_1983}. We normalised the injection function for each outflow component with the condition:
\begin{equation}
\int_{E_{\rm min}}^{E_{\rm max}} E\,Q_{p(e)}(E)\,{\rm d}E = L_{{\rm inj, NT}p(e)}\,,
\label{Eq:Norm_Qinj}
\end{equation}
with $E_{{\rm min,}p} = 1$~GeV for protons and $E_{{\rm min,}e} = 2m_ec^2$ for electrons; only very high $E_{{\rm min,}p,e}$ values would affect significantly the results. As for the maximum energy, $E_{\rm max}$, we took into account the rate at which particles gain energy:
\begin{equation}
    \dot{E} = \frac{E}{t_{\rm acc}} = \eta_{\rm acc}cq_eB_1,
    \label{Eq:Edot_acc}
\end{equation}
where $q_e$ is the electron charge, and $\eta_{\rm acc} = \left(v_{\rm out}/c\right)^2/\left(2\pi\right)$ is the acceleration efficiency. It is worth noting that we considered $P_1$ and the outflow velocity $v_{\rm out}$ (instead of $v_1$) in Eq.~\ref{Eq:Edot_acc}, because the pressure $P_1$ remains constant throughout the mixing process, and therefore $B_1$ is also constant between the shock and the mass-loaded tube, while the acceleration efficiency depends on the shock velocity. Equation~\ref{Eq:Edot_acc} defines the acceleration timescale:
\begin{equation}
    t_{\rm acc} = \frac{1}{\eta_{\rm acc}}\frac{E}{c\,q_e\,B_1} \quad {\rm s}\,,
    \label{Eq:t_acc}
\end{equation}
 setting the maximum energy by the intersection between the acceleration and total loss timescales, which takes into account both escape and radiative losses:
\begin{equation}
    t_{\rm loss} = \frac{1}{t_{\rm esc}^{-1}+t_{\rm cool}^{-1}} \quad {\rm s}\,.
    \label{Eq:t_loss}
\end{equation}
In Eq.~(\ref{Eq:t_loss}), the escape timescale is $t_{\rm esc} = h_1/v_1$, while $t_{\rm cool}$ depends on the cooling mechanisms of each particle species. Protons cool via adiabatic losses and proton-proton ({\it pp}) collisions. The adiabatic loss timescale depends on the evolution of the shocked outflow density in the tube \citep{del_Palacio_2018}:
\begin{equation}
    t_{\rm adi} = \frac{3}{v}\frac{{\rm d}z}{{\rm d}\left(-\ln{\left(\rho\right)}\right)}\quad {\rm s}\,. 
\end{equation}
Regarding {\it pp} collisions, the target density for the mechanism within the shocked outflow is low ($n \sim 1$~cm$^{-3}$). However, protons may also interact with material surrounding the outflow (i.e. the shocked or unshocked core). Since this dense wall is difficult to characterise, due to the possible development of overdense and unstable regions, we adopted a phenomenological approach. We assumed that protons interact with material of density $n = 5 \times n_{\rm core} = 5 \times 10^5$~cm$^{3}$ and computed the resulting pion emissivity and $\gamma$-ray emission. These predictions should be regarded as upper limits, as only a fraction of the relativistic protons may interact with the dense walls before the outflow disruption. The cooling timescale of {\it pp} losses is given in terms of the target density, $n$:
\begin{equation}
    t_{pp} \approx \frac{1}{c\,n\,K_{pp}\,z_{pp}\,\sigma_{pp}\left(E_p\right)}\Theta\left(E_p - 1.22~{\rm GeV}\right) \quad {\rm s}\,,
    \label{Eq:t_pp}
\end{equation}
with $\Theta(E)$ the Heaviside function, $K_{pp} \approx 0.5$, $z_{pp} = 2.17$ \citep{Padovani_2018}, and we used the standard cross-section from \cite{Kelner_2006}.

On the other hand, electrons cool via adiabatic losses, synchrotron, relativistic bremsstrahlung, and inverse Compton (IC) interactions with the dust IR photon field, although the last of these is not relevant in this scenario. The timescale of the former is $t_{\rm syn} \approx 4 \times 10^{13}\,B^{-2}\,\left(E_e/{\rm eV}\right)^{-1}$~s. Finally, relativistic bremsstrahlung depends on the density of proton targets. Once again, we set the target density phenomenologically at $n = 5\times 10^5$~cm$^{-3}$, as made for $pp$ collisions. The timescale of this cooling process is $t_{\rm Br} \approx 10^{15}/\left(z_{\rm Br}\,n\right)$~s, with $z_{\rm Br} = 2.24$ \citep{Padovani_2018}. The formulae of cooling via IC, which is negligible, can be found in \cite{del_Palacio_2018} and references therein.

We employed the approximation from \cite{del_Palacio_2018} to calculate the steady-state particle distribution at the injection cell considering escape and radiative losses:
\begin{equation}
    N_1(E) \approx Q(E)\,\times\,\min{\left(t_{{\rm cell},1},t_{{\rm cool},1}\right)},
    \label{Eq:N_1}
\end{equation}
with $t_{\rm cell} = {\rm d}z(l)/v(l)$ being the cell advection time.

In the following cells, we considered conservation of the number of particles along with energy losses. These constraints set the evolution of the maximum energy and particle energy distributions cell by cell as:
\begin{equation}
    N(l,E') = N(l-1,E)\,\frac{|\dot{E}(E,l)|}{|\dot{E'}(E',l)|}\,\frac{t_{\rm cell}(l)}{t_{\rm cell}(l-1)},
    \label{Eq:N_e}
\end{equation}
where $\dot{E}(E,l) = -E/t_{\rm cool}(E, l)$ is the energy loss rate, and $E' \approx E + \dot{E}\,t_{\rm cell}$. Finally, we notice that our model does not consider the putative leptonic emission originated at the base of the outflows, prior to the deflection \citep{Barkov_2012,Fender_2013,Tsuna_2019,Scarcella_2021}. This contribution could enhance the non-thermal emission of the source, but is expected to be sub-dominant.

%--------------------------------------------------------------------
%====================================================================

\subsubsection{Shocked medium}\label{Subsubsec:shocked_med}

The forward shock that propagates through the medium is radiative since the shocked gas convection velocity is $\sim v_{\rm IBH}$, and the gas suffers thermal cooling before convecting away along the tube. We employed Rankine-Hugoniot jump conditions to calculate the thermodynamic quantities right after the shock. The thermal spectrum is largely dependent on the temperature, which in that region is proportional to $v_{\rm IBH}^2\sin^2{\left(\alpha\right)}$. The charged particles of the ionised gas, also dependent on the metallicity, emit post-recombination lines and thermal bremsstrahlung radiation in the continuum. In particular, for solar abundances and velocities of $v_{\rm IBH} = 30$~km s$^{-1}$, cooling is dominated by line emission, such as H$_\alpha$ and H$_\beta$, but they are absorbed inside the core and re-emitted in the IR.

In order to calculate the ionisation fraction $x$ as a function of temperature, we employed the Saha equation for hydrogen:
\begin{equation}
    \frac{x^2}{1-x} = \frac{1}{n} \left(\frac{2 m_e k T(l)}{h^2} \right)^{3/2} \exp{\left(-\frac{E_{\rm ion}}{kT(l)}\right)}\,,
    \label{Eq:Saha}
\end{equation}
where $T(l)$ is the shocked medium temperature right after the shock, $n$ is the particle number density, and $E_{\rm ion} = 13.6$~eV is the hydrogen ionisation energy. Once we computed the number density of ions and electrons, we calculated the intrinsic H$_\alpha$ emission adapting the expressions given in \cite{Mackey_2013, Gvaramadze_2018} and  \cite{Martinez_2023}:
\begin{equation}
    L_{{\rm H}_\alpha} = 1.72\,\pi\times 10^{11}\,\kappa\left(l\right)\,\left(\frac{{\mathscr V}\left(l\right)}{{\rm cm}^3}\right) \quad {\rm erg\,s^{-1}}\,,
    \label{Eq:L_Halpha}
\end{equation}
with $\mathscr{V}(l) = t_{\rm th}(l)\,\left(v_{\rm IBH}/\xi\right)\,\sin{\left(\alpha(l)\right)}\,S(l)$ the volume of the shocked medium at the $l-$th cell, where $t_{\rm th}$ is the gas cooling timescale \cite[see the expression for $t_{\rm th}$ in][]{Stevens_1992,Myasnikov_1998}, and $\xi$ the compression factor. On the other hand, $\kappa\left(l\right)$ is dependent on the temperature and number density of electrons and ions:
\begin{equation}
    \kappa\left(l\right) = 2.85 \times 10^{-33}\,\left(\frac{T\left(l\right)}{{\rm K}}\right)^{-0.9}\,\left(\frac{n_{\rm e}\left(l\right)}{{\rm cm}^{-3}}\right)\,\left(\frac{n_{\rm ion}\left(l\right)}{{\rm cm}^{-3}}\right)\,.
    \label{Eq:kappa_alpha}
\end{equation}
Then, we estimated the intrinsic H$_\beta$ luminosity as $L_{{\rm H}_\beta} \approx L_{{\rm H}_\alpha}/3$ \citep{Hummer_1987}.

Finally, we calculated the thermal bremsstrahlung radiation. To calculate the continuum spectrum we used the formula given by \cite{Rybicky_Lightman}:
\begin{equation}
    L_{\rm ff}(\epsilon,l) = 6.8\times 10^{-38} \, \frac{n_{\rm e}(l)\,n_\mathrm{ion}(l)\,\mathscr{V}(l)}{\sqrt{T(l)}\,h} \, e^{-\epsilon/kT(l)} \,\bar{g}_\mathrm{ff} \quad {\rm erg\,s^{-1}},
    \label{Eq:L_ff}
\end{equation}
where $\bar{g}_{ff}$ is the velocity averaged Gaunt Factor, the values of which are tabulated in \cite{Van_Hoof_2014}. Lastly, we notice that the thermal emission of the shocked outflow within the tube is negligible due to the low densities.

%--------------------------------------------------------------------
%====================================================================

\subsection{Injection into the molecular cloud}\label{Subsec:MC_core}

After the tube gets disrupted due to instability growth, its material mixes with the material of the surrounding medium, effectively spreading its content in a wide region. This results in the injection of high-energy protons and electrons into the core of the molecular cloud. These particles diffuse through the cloud and can interact with radiation fields and matter, emitting non-thermal radiation.

To estimate the emission of the non-thermal particles diffusing in the medium, we employed one-zone models for both the core and the rest of the molecular cloud. We considered the energy distribution of the particles emerging from the tube, which is equivalent to assuming an injection distribution that follows a power-law with spectral index 2 and an exponential cutoff as in Eq.~(\ref{Eq:Q_inj}). We calculated the normalisation constant of the injection function with the condition:
\begin{equation}
    \int_{E_{{\rm min},p(e)}}^{{E_{{\rm max},p(e)}}} E\,Q_{{\rm inj},p(e)}^{\rm core}(E)\,{\rm d}E = L_{{\rm inj, NT},p(e)}^{\rm core},
    \label{Eq:L_inj_MCC}
\end{equation}
where $ L_{{\rm inj, NT},p(e)}^{\rm core}$ is the luminosity injected into protons(electrons) in the core, $E_{{\rm min},p} = 1$~GeV, $E_{{\rm min},e} = 2m_ec^2$, and $E_{\rm max}$ is the maximum particle energy at the last cell of the tube. To compute the injected luminosity, we considered the injection by both the forward and the rear outflow, and summed the contributions for both relativistic protons and electrons separately. Each component injects a luminosity into each particle type in the core of:
\begin{equation}
    L_{{\rm inj, NT},p(e)}^{\rm core} = L_{{\rm NT},p(e)}(l_\mathrm{max}) = \frac{1}{t_{\rm cell}(l_\mathrm{max})} \int E\,N_{p(e)}\left(E,l_\mathrm{max}\right)\,{\rm d}E \,.
    \label{Eq:L_inj_core}
\end{equation}
Then, we approximated the steady-state particle distribution as:
\begin{equation}
    N_{p(e)}^{\rm core}(E) \approx Q_{{\rm inj},p(e)}^{\rm core}(E)\times \,t_{{\rm tot},p(e)}(E),
    \label{Eq:N_MCC(E)}
\end{equation}
where $t_{\rm tot}^{\rm -1} = \left(t_{\rm esc}^{\rm -1} + t_{\rm cool}^{\rm -1}\right)$ is the total cooling timescale. In turn $t_{\rm esc}^{\rm -1} = \left(t_{\rm cross}^{\rm -1} + t_{\rm diff}^{\rm -1}\right)$ is the escape timescale, whereas $t_{\rm cool}$ accounts for radiative cooling, as particles do not suffer adiabatic losses because the cloud medium was considered static. We assumed that particles diffuse under the Bohm regime, while $t_{\rm cross}$ is the IBH crossing timescale through the core, the time at which the system accelerates non-thermal particles efficiently while accreting material from the dense medium. This timescale is given by $t_{\rm cross} = R_{\rm core}/v_{\rm IBH}$, with $R_{\rm core} \sim 0.1$~pc being the size of the core. We note that, although the diffusion regime is not well known, the dense medium and the perturbation by the disrupting flow favour slow diffusion. Adopting Bohm diffusion is thus useful to provide a plausible but optimistic assessment of the emission of the diffusing particles. We considered that protons are cooled by $pp$ interactions that give rise to emission at high energies. Electrons are cooled by relativistic bremsstrahlung, and by synchrotron to a lesser extent, while IC losses are negligible. To calculate the synchrotron spectrum we set a magnetic field of 100~$\mu$G in the cloud core \citep[it might be 10~$\mu$G in the rest of the molecular cloud; e.g.][]{Crutcher_1991}.

Once the non-thermal particles have escaped from the core, they are injected into the rest of the molecular cloud. We characterised this region as a homogeneous sphere with a density of $n_{\rm MC} = 10^2~$cm$^{-3}$ and size $R_{\rm MC} = 1$~pc. We employed a second one-zone model with a power law injection function employing Eqs.~(\ref{Eq:Q_inj}) and (\ref{Eq:L_inj_MCC}), but with the injection luminosity $L_{\rm inj, NT}^{\rm MC} = L_{\rm inj, NT}^{\rm core} - L_{\rm rad}^{\rm core}$, where $L_{\rm rad}^{\rm core}$ is the luminosity radiated within the core. The steady-state particle distribution is:
\begin{equation}
    N_{p(e)}^{\rm MC}(E) \approx Q_{{\rm inj,}p(e)}^{\rm MC}(E)\times \min{\left( t_{\rm cross}, t_{\rm diff}\right)}\,,
    \label{Eq:N_MC(E)}
\end{equation}
where we considered Bohm diffusion once again. We note that both cooling and advection are negligible compared to the diffusion and crossing timescales. Finally, we calculated the corresponding emission from this region following an analogous approach to that employed for the core.

%--------------------------------------------------------------------
%====================================================================
\section{Results and discussion}\label{Sec:Results}
%--------------------------------------------------------------------
%--------------------------------------------------------------------
\subsection{Accretion emission}\label{subsec:res_acc_disk}

\begin{figure*}[t]
    \centering
    \includegraphics[width=0.49\textwidth]{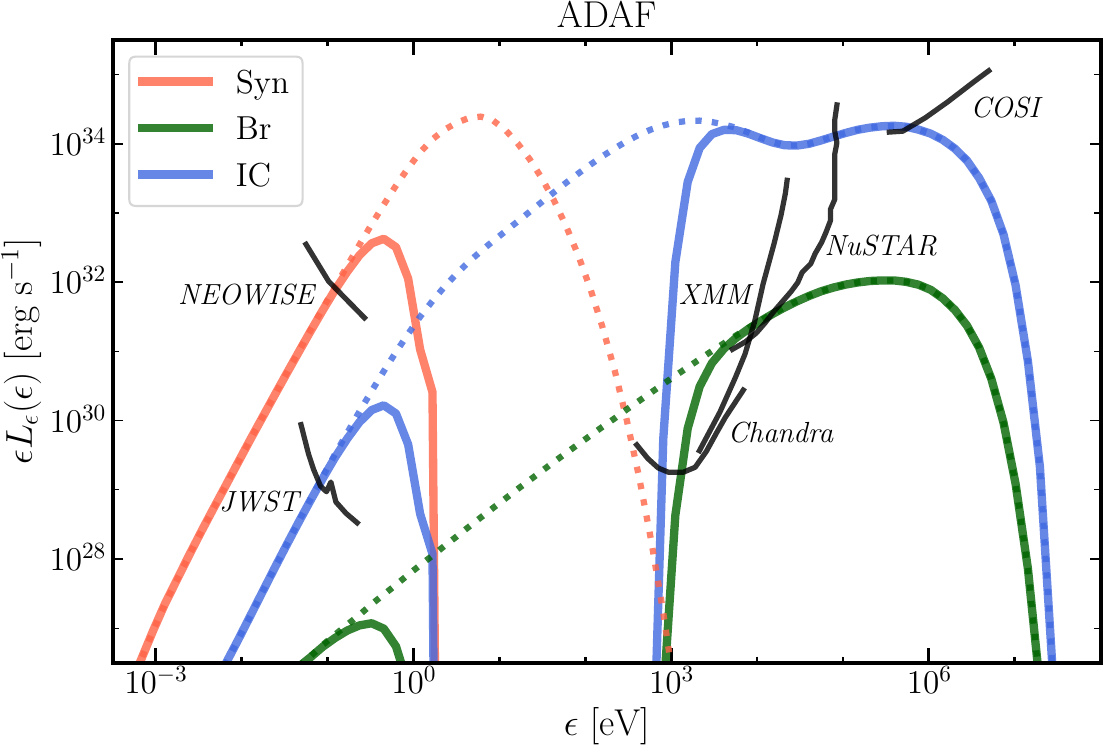}
    \includegraphics[width=0.49\textwidth]{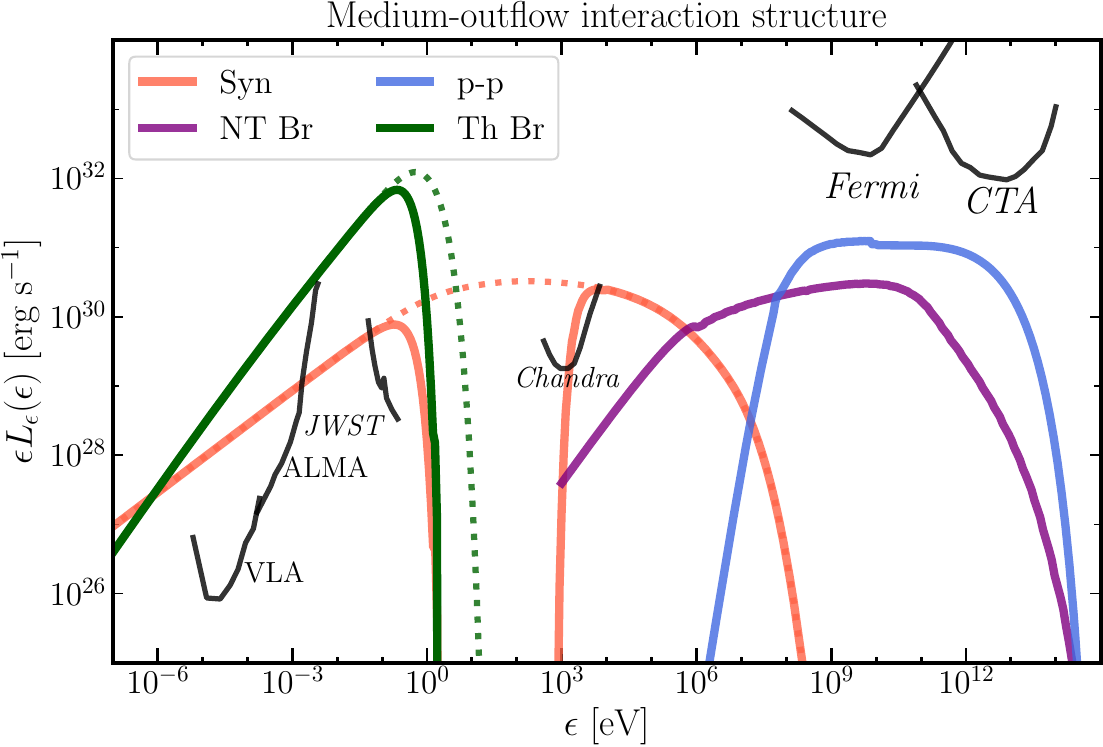}
    \includegraphics[width=0.49\textwidth]{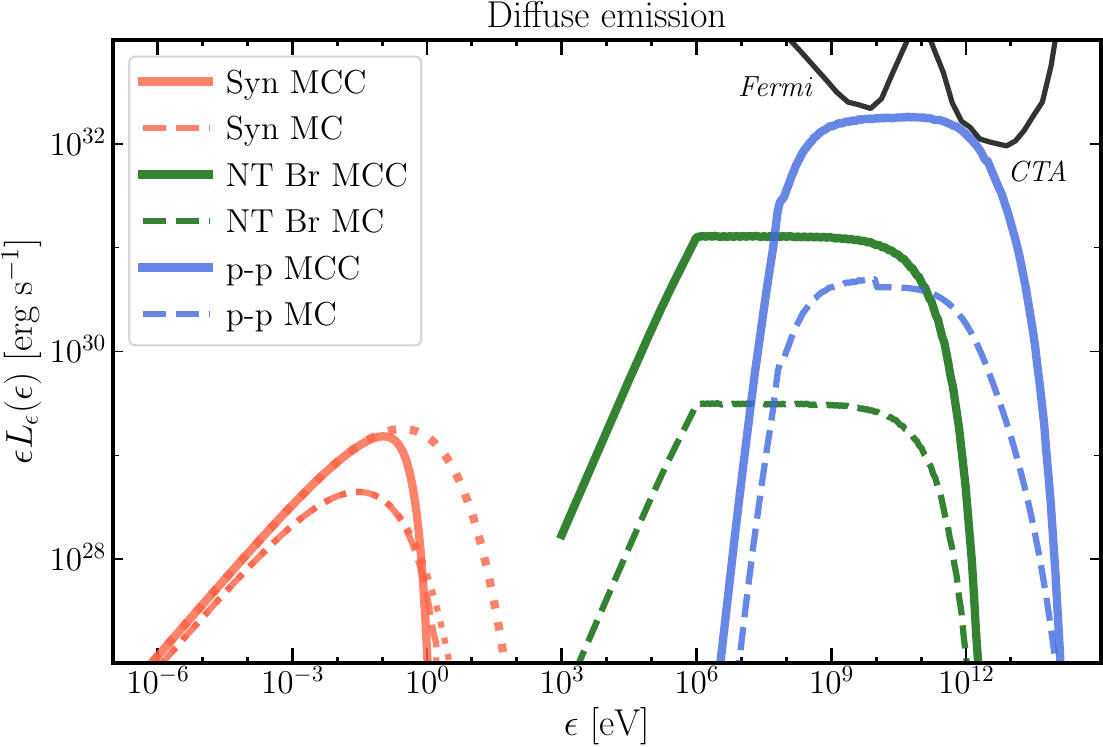}
    \includegraphics[width=0.49\textwidth]{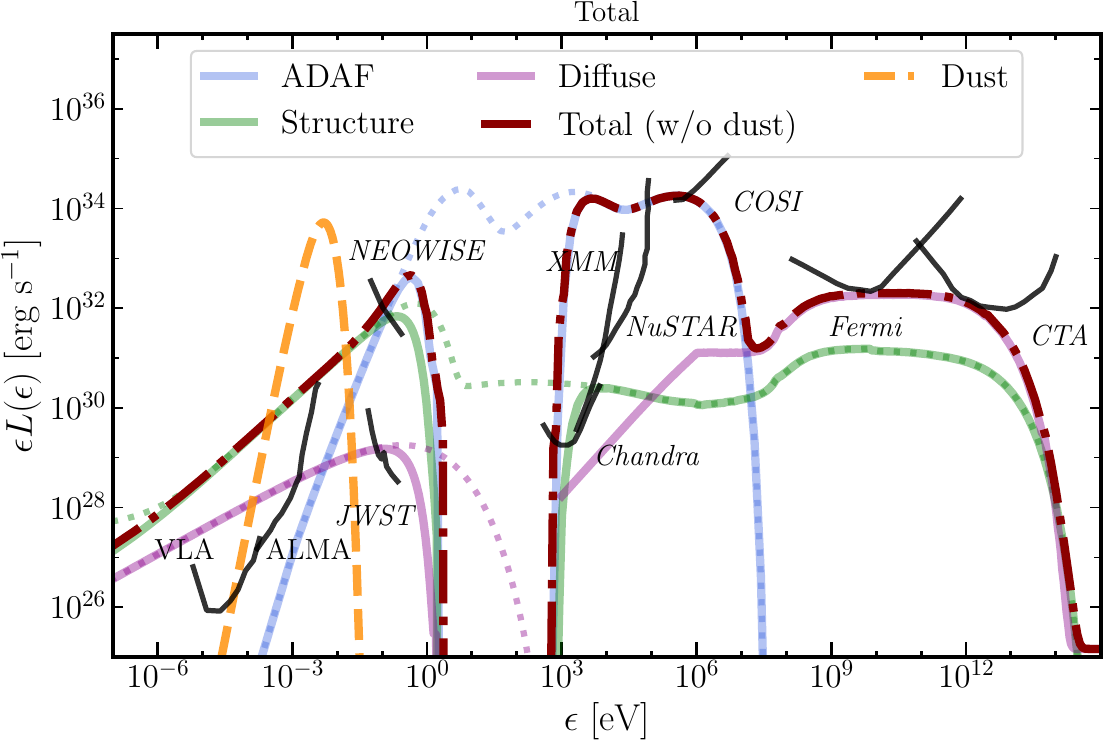}
    \caption{Spectrum of an IBH inside a molecular cloud core for the parameters listed in Table~\ref{table:IBH_parameters}. The dotted and solid lines represent the absorbed and unabsorbed spectra, respectively. The VLA sensitivity curve corresponds to a total observation time of 1~h, while the ALMA sensitivity curve corresponds to an on-sourve time of 1~h. We extracted the 10~ks MIRI sensitivity curve from the instrument user documentation (\url{https://jwst-docs.stsci.edu/jwst-mid-infrared-instrument/miri-performance/miri-sensitivity}). The {\it NEOWISE} sensitivity is taken from \cite{NEOWISE}. The {\it Chandra}, {\it XMM-Newton} and {\it NuSTAR} sensitivity curves correspond to observation times of 100~ks, and were taken from \url{https://chandra.harvard.edu/about/specs.html}, \cite{Ferrando_2002} and \cite{NuStar}, respectively. The COSI sensitivity corresponds to a 2 yr survey time \citep{COSI_2023}. The {\it Fermi} sensitivity curve corresponds to a 10~yr observation for sources with Galactic latitude $|b| < 30\degree$ (see \url{https://www.slac.stanford.edu/exp/glast/groups/canda/lat_Performance.htm}). The CTA sensitivity curve corresponds to a 50~h observation time \citep{CTA}. We assumed a distance to the source of 2~kpc. {\it Top left panel:} accretion emission. The hot electrons emit via synchrotron and bremsstrahlung, and also Comptonise the low-energy photons. {\it Top right panel:} shocked medium thermal emission and shocked outflow non-thermal emission. {\it Bottom left panel:} non-thermal emission from protons and electrons diffusing through the core (solid lines) and the outer regions of the molecular cloud (dashed lines). {\it Bottom right panel:} Total spectrum, including the emission from the molecular cloud dust.}
    \label{Fig:SED_total}
\end{figure*}

In the upper left panel of Fig.~\ref{Fig:SED_total} we show the spectrum produced by the hot electrons of the ADAF, which exhibits three distinct peaks. The first peak is the result of synchrotron emission, which reaches its maximum in the optical range. However, the molecular cloud core absorbs radiation within the range between the near IR and soft X-rays. For this reason, the mid-IR band is the most promising for detecting synchrotron radiation with telescopes such as the Mid-IR Instrument (MIRI) of the \textit{James Webb Space Telescope} or the {\it Near-Earth Object Wide-field Infrared Survey Explorer (NEOWISE)}. Moreover, synchrotron photons can undergo multiple Comptonisation events, resulting in a second and a third peak in the spectrum. Taking into account that the cloud is transparent above $\sim 1$~keV, this emission could be detected with X-ray observatories such as {\it Chandra}, {\it XMM-Newton} or {\it NuSTAR} even if the IBH were much further away than the considered distance of 2~kpc. Moreover, the accretion ADAF spectrum would be observable with the forthcoming {\it Compton Spectrometer and Imager} (COSI) around $E = 1$~MeV. We highlight that the expected spectrum would be analogous to that of a microquasar with a similar accretion state \cite[e.g.][]{esi98}. The identification of an IBH is thus contingent upon the analysis of the light curve, which should not exhibit any feature characteristic of a binary system. It is worth noting that the integrated unabsorbed luminosity between 100 eV and 1 MeV is $L_{\rm X, ADAF} \sim 2\times 10^{35}$~erg s$^{-1}$, which is close to the value of $L_{\rm X, ADAF} \sim 3\times 10^{35}$~erg s$^{-1}$ predicted by Eq.~(\ref{Eq:L_X_ADAF}).

Finally, we also addressed alternative scenarios in the Appendix~\ref{Appendix_discs}, such as an IBH within the Galactic disc (and not crossing though a molecular cloud) or the transient formation of a thin accretion disc inside a cloud core. According to our results, an IBH in a medium with Galactic disc densities would not be detectable. On the other hand, the standard disc in a dense medium would be observable in the IR and in X-rays.

%--------------------------------------------------------------------
\subsection{Deflected outflow and shocked medium}\label{subsec:res_out}

The shocked medium produces intrinsically bright H$_{\alpha}$ and H$_\beta$ emission for the parameters considered in Table~\ref{table:IBH_parameters}. From a total luminosity injected into the shocked medium of $L_{\rm inj}^{\rm SMED} \sim 2 \times 10^{33}$~erg s$^{-1}$, a luminosity of $L_{{\rm H}_\alpha} \sim 10^{33}$~\ergs and $L_{{\rm H}_\beta} \sim 3 \times 10^{32}$~\ergs are radiated through H$_\alpha$ and H$_\beta$ lines, respectively. However, line emission is absorbed within the dense core and re-emitted in the IR, accounting for $\approx 30\%$ of the dust spectrum shown in the bottom right panel of Fig.~\ref{Fig:SED_total}. In the continuum, the shocked medium emits a luminosity of $L_{\rm ff} \sim 4 \times 10^{32}$~\ergs from radio to optical wavelengths.

The conditions at the outflow deflection point, where a strong shock should develop, can be adequate for efficient particle acceleration. The high outflow velocity and a magnetic field of the order of mG can lead to a high acceleration efficiency by diffusive shock acceleration, whose timescale is given in Eq.~(\ref{Eq:t_acc}). 

\begin{figure}[t]
    \centering
    \includegraphics[width=0.49\textwidth]{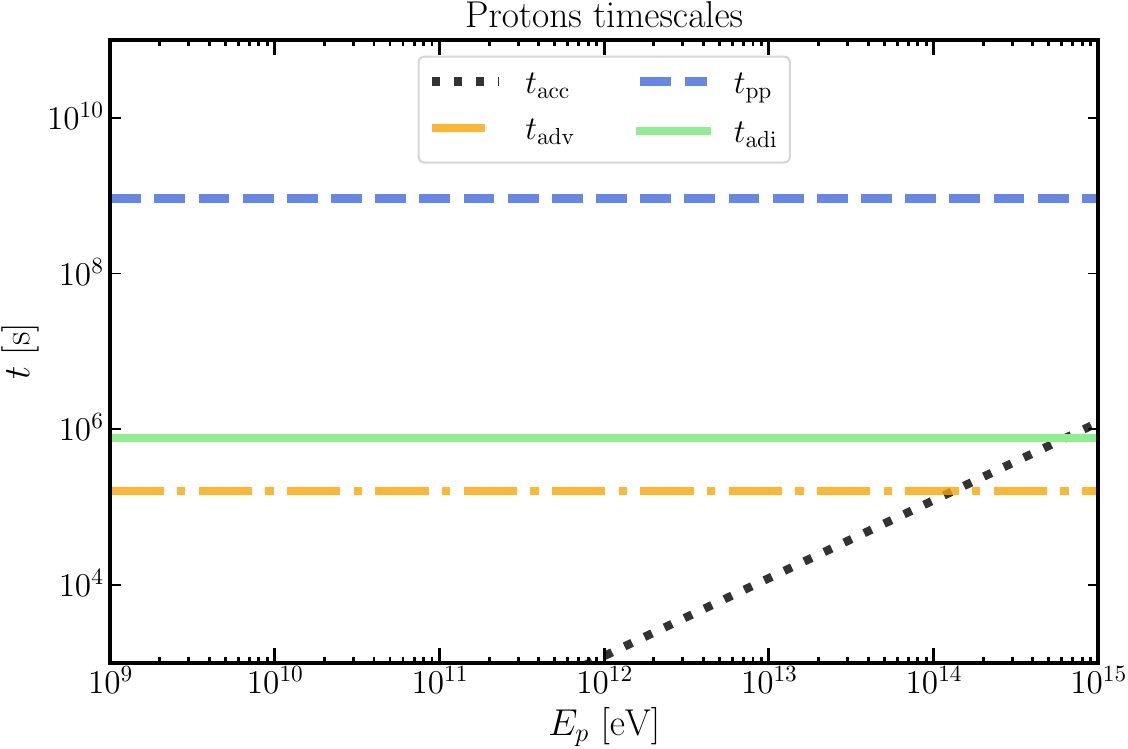}
    \label{fig:tiempos_p}
\\
    \centering
    \includegraphics[width=0.49\textwidth]{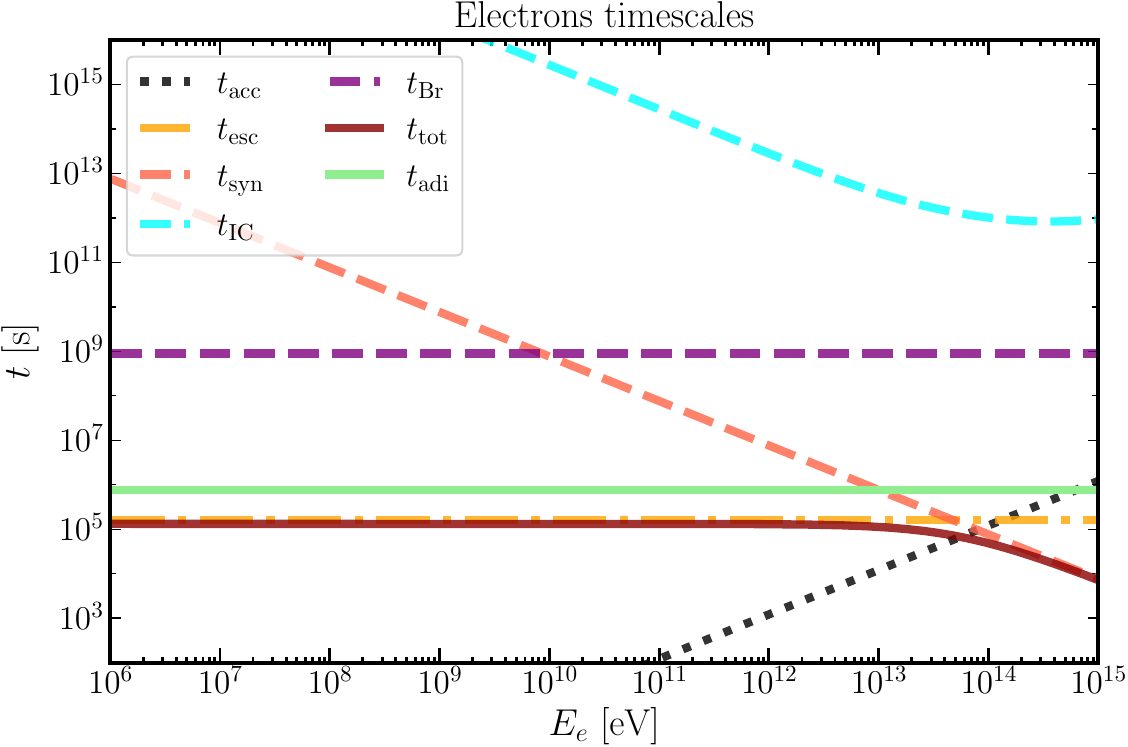}
    \caption{Characteristic timescales for protons (top panel) and electrons (bottom panel) at the deflection point of the forward outflow. Electrons reach maximum energies of $E_{{\rm max},e} \sim 60$~TeV, while protons reach $E_{{\rm max},p} \sim 100$~TeV. Protons escape from the structure without radiating significantly. Along the tube, the most energetic electrons suffer significant synchrotron cooling.}
    \label{fig:tiempos_e}
\end{figure}

\begin{figure}[t]
    \centering
    \includegraphics[width=0.49\textwidth]{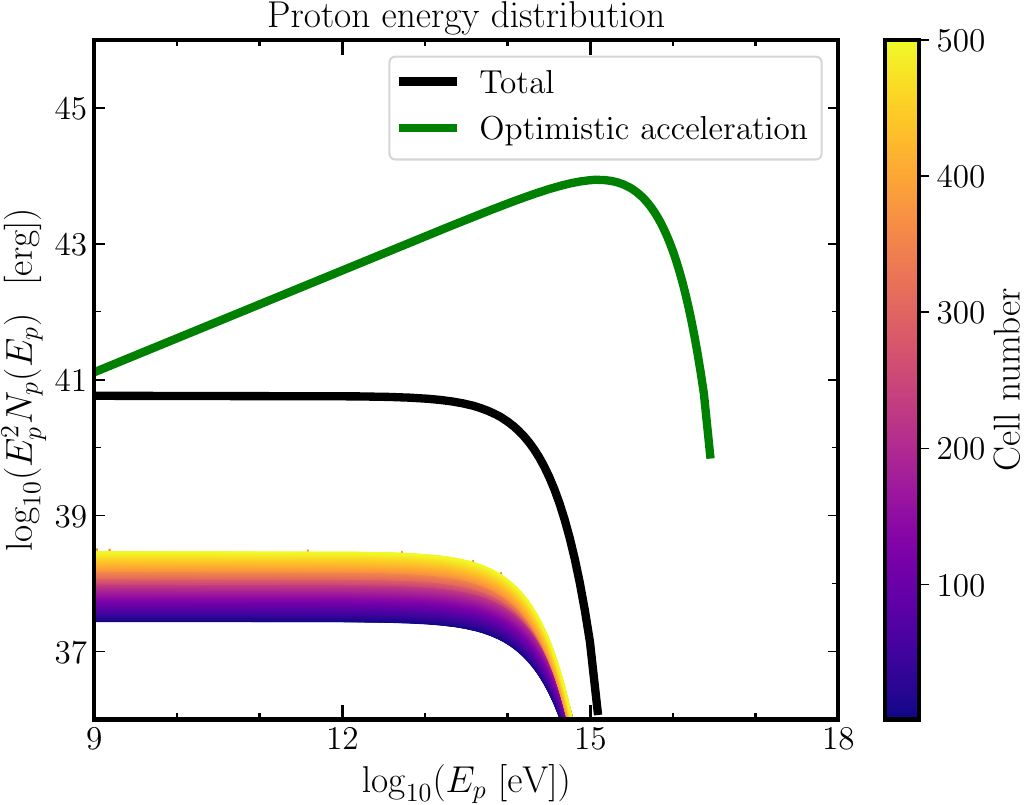}
    \label{fig:distp}
\\
    \centering
    \includegraphics[width=0.49\textwidth]{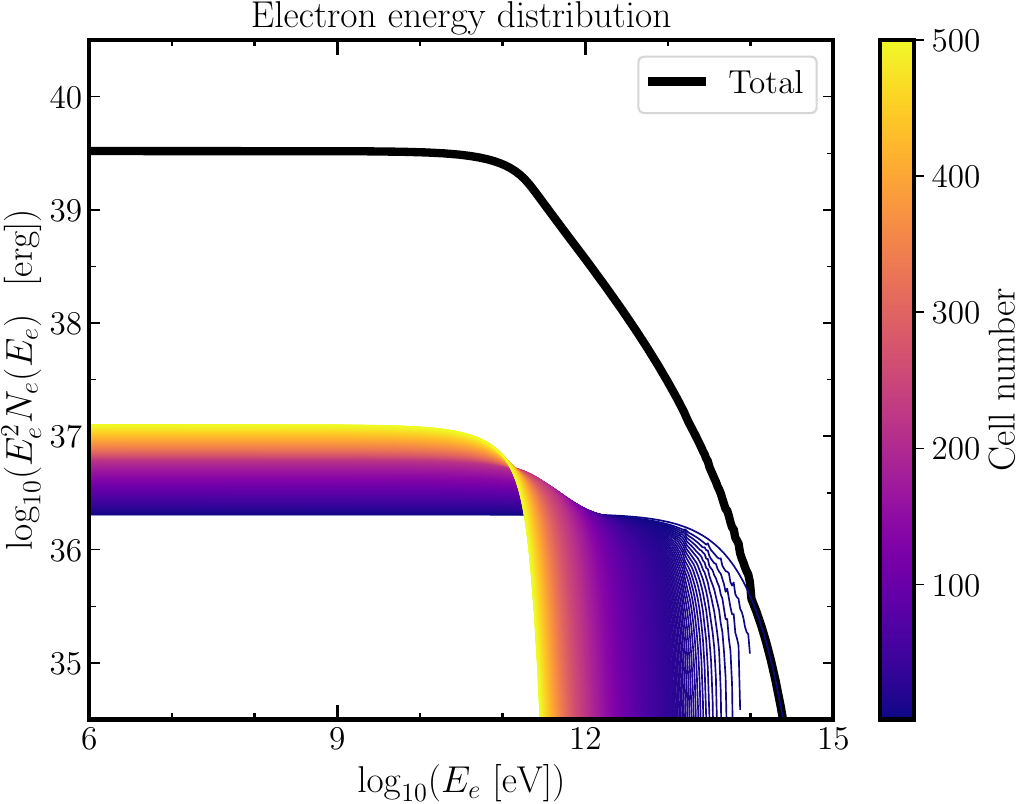}
    \caption{ Relativistic particle energy distributions for protons (top panel) and electrons (bottom panel). The coloured scale shows the distribution at each cell. The black line represents the total energy distribution (i.e. the sum over all cells). The green line corresponds to an alternative scenario with optimistic acceleration and injection, where the maximum energy is given by $E_{\rm max} = q_e\,\sqrt{4\,\left(0.5L_{\rm out}\right)/c}$, the luminosity injected into protons is $L_{\rm inj}^{p} = 0.1\,L_{\rm out}$, and with a harder spectral index $p>-2$. The optimistic approach is not applicable for electrons due to synchrotron losses at energies $E \gtrsim 120$~GeV.}
    \label{fig:diste}
\end{figure}

Figures~\ref{fig:tiempos_e} and \ref{fig:diste} illustrate the characteristic timescales and energy distributions for both protons and electrons and show that non-thermal particles can be accelerated up to very-high energies. Assuming a magnetic field parameter of $\eta_{\rm B} = 0.1$, electrons reach energies of $E_{{\rm max},e} \sim 60$~TeV, while protons reach energies of $E_{{\rm max},p} \sim 100$~TeV. A larger value of $\eta_{\rm B}$ and of $\eta_{\rm acc}$ would lead to more efficient acceleration. However, $\eta_{\rm B}$ is restrained below unity, as otherwise the fluid would become incompressible and diffusive shock acceleration would not take place. The considered parameters lead to a total injection luminosity into protons of $L_{\rm inj, p}^{\rm out} \approx 2\times 10^{34}$~erg\,s$^{-1}$, and into electrons of $L_{\rm inj, e}^{\rm out} \approx 2 \times 10^{33}$~erg\,s$^{-1}$. We recall that $\eta_{\rm NT}=0.1$, so there is room for a higher non-thermal p/e power.

The system can produce a significant non-thermal population of very energetic particles, but their emission luminosity depends on the magnetic, matter and radiation fields available. Synchrotron and relativistic bremsstrahlung emission are the most significant emission mechanisms for electrons. Through these processes, we predict a luminosity of $L_{\rm syn} \sim 4 \times 10^{31}$~erg s$^{-1}$ from radio to X-rays, and $L_{\rm Br} \sim 3 \times 10^{31}$~erg s$^{-1}$ from X-rays up to $\gamma$-rays, respectively. Regarding protons, we predict a putative luminosity of $L_{pp} \sim 2 \times 10^{32}$~erg s$^{-1}$ at $\gamma$-rays from the interaction structure, which would not be detectable with current nor forthcoming instrumentation for the parameters assumed in Table~\ref{table:IBH_parameters}.

In the upper right panel of Fig.~\ref{Fig:SED_total} we show both the free--free spectrum of the shocked medium along with the non-thermal spectrum of the deflected outflow. Although a dense core is optically thin at low frequencies, its total radiation dominates the millimetre band.  However, the shocks are strong near the outflow deflection points, implying surface brightness higher than that of the cloud at frequencies $\nu \lesssim 300$~GHz. For this reason, we simulated emission maps at different frequencies to ascertain the detectability of the structure with the available instruments. Given that the bright region of the structure extends to less than 10" for the considered distance of 2~kpc, an instrument with adequate angular resolution is required to resolve the emission of this structure from that of the accretion.

\begin{figure*}[ht]
    \centering
    \includegraphics[width=0.9\textwidth]{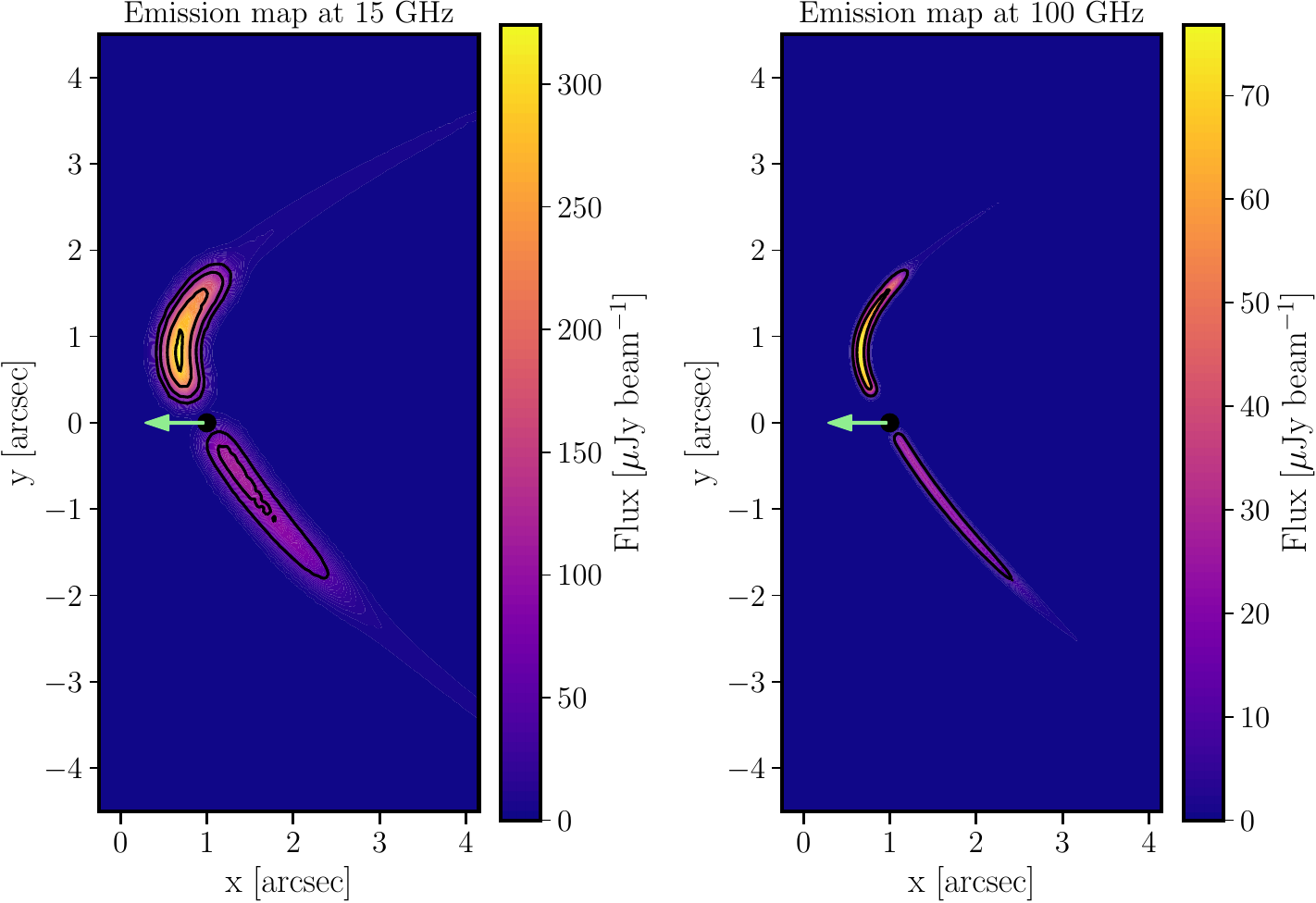}
    \caption{Simulated emission maps at 15~GHz (left) and at 100~GHz (right panel) with beam sizes of $0.13" \times 0.13"$ and $0.042" \times 0.042"$, respectively. The forward interaction structure is a bright arc. The black dot shows the position of the IBH, and the green left-arrow shows its direction of motion. The black contour levels are at 50, 100, 200 and 300~{\rm $\mu$}Jy beam$^{-1}$ in the left panel, and at 10 and 50~{\rm $\mu$}Jy beam$^{-1}$ in the right panel.}
    \label{fig:emission_maps}
\end{figure*}

First, we considered the Very Large Array\footnote{\url{https://obs.vla.nrao.edu/ect/}} (VLA) radio interferometer as a candidate instrument to detect radio emission from the shocked medium. In the A array configuration, this interferometer has a beam size of $\approx$ 0.13" at 15~GHz. In the left panel of Fig.~\ref{fig:emission_maps} we show the emission map at 15~GHz including the shocked medium and outflow considering a line of sight perpendicular to the structure. We predict peak flux densities of $S_\nu \sim 300$~$\mu$Jy beam$^{-1}$, above the detection level of the instrument for an on-source observation time\footnote{\url{https://science.nrao.edu/facilities/vla/docs/manuals/oss/performance/resolution}} of 1~h. At these low frequencies, although the free-free emission is dominant, the contribution of synchrotron radiation from the shocked outflows is also noticeable, since the synchrotron flux is $S_\nu \propto \nu^{-0.5}$ for an electron energy distribution $N_e(E) \propto E_e^{-2}$.

Secondly, we analysed the outflow-medium interaction structure detectability with the Atacama Large millimetre Array\footnote{\url{https://almascience.eso.org/documents-and-tools/cycle10/alma-technical-handbook}} (ALMA). This instrument achieves a maximum angular resolution of 42~mas at $\nu = 100$~GHz with a largest angular scale of $\approx 0.5$", filtering the dust emission, which has an angular scale of 10" and a surface brightness of $\approx 10$~$\mu$Jy~as$^{-2}$. In the bottom panel of Fig.~\ref{fig:emission_maps} we show the emission map at 100 GHz, that is dominated by the emission from the shocked medium, and where we obtained peak flux densities of about 85~$\mu$Jy~beam$^{-1}$. These values are above the sensitivity of the instrument, which is $\approx 10$~$\mu$Jy~beam$^{-1}$ for 1~h on-source time\footnote{\url{https://almascience.eso.org/proposing/sensitivity-calculator}}. Therefore, the deflected outflows would be detectable with ALMA as well.

We note that the synchrotron spectrum reaches energies around 1~MeV due to the presence of electrons with energies around $E_e \sim 60$~TeV. Moreover, the synchrotron radiation fluxes are above the detectability threshold of MIRI at IR and of {\it Chandra} at X-rays. Nevertheless, the instruments do not have adequate angular resolutions to resolve the shocked outflow emission from that of accretion. Then, the interaction structure non-thermal emission would only be detectable at these energies in a scenario with no ADAF formation, as the scenarios proposed by \citealt{Barkov_2012_b} and \citealt{Barkov_2012}.

Finally, we also accounted for the structure emission if mixing at the deflection points did not occur. Under this assumption, Rankine-Hugoniot conditions lead to a density $\rho_1 = 4\,\rho_{\rm out}$ and the advection velocity becomes $v_1 = \left(3/4\right)\,v_{\rm out}$. This implies that escape would dominate electron losses even at energies > 1~TeV, and that synchrotron luminosity would decrease by a factor of $\sim$ 4. On the other hand, the considered value of $v_1$ yields a maximum Doppler boosting of $\delta_1^3/\gamma_1 = 4.77$, with:
\begin{equation}
    \delta_1\left(\theta_{\rm obs}=0\right) = \frac{1}{\gamma_1\left(1-\beta_1\right)} \approx 1.73\,,
\end{equation}
with $\theta_{\rm obs}$ being the angle between the line of sight and the outflow direction. This indicates that the use of a non-relativistic flow model introduces only relatively small errors in the results at our level of approximation, and that our main conclusions do not strongly depend on the mixing occurrence.

We conclude that the medium-outflow interaction structure of an IBH in a dense core could be detected with high angular resolution instruments. In particular, we highlight the possible detectability in radio and in the millimetre range, and that the structure could be resolved with VLA and with ALMA. In addition, the forthcoming radio interferometers, such as the Square Kilometre Array\footnote{\url{https://www.skao.int/en}}, or the Next Generation Very Large Array\footnote{\url{https://ngvla.nrao.edu/}} could detect fainter interaction structures.

%--------------------------------------------------------------------
\subsection{Core and cloud non-thermal emission}\label{subsec:res_dif}

Protons escape from the tube mostly affected by adiabatic losses, and consequently the hadronic cosmic ray (CR) luminosity injected into the molecular cloud core is $L_{\rm inj,NT}^{\rm core} \sim 10^{34}$\ergs. Diffusion determines the residence time of particles in the core, but for the lowest energy protons, the relevant timescale is rather the IBH core crossing time, $t_{\rm cross} \sim 1000$~yr. Considering a spherical core of radius $R_{\rm core} = 0.1$~pc, this yields a CR energy density of $u_{\rm CR} \sim 1000$~eV\,cm$^{-3}$, exceeding by far the $\sim 1$~eV\,cm$^{-3}$ corresponding that of the diffuse Galactic CRs \citep{Greiner_2015}.

In the lower right panel of Fig.~\ref{Fig:SED_total} we show the spectrum of non-thermal electrons and protons diffusing both within the core and throughout the outer regions of the molecular cloud. The hadronic emission within the dense core dominates the spectrum between 10~MeV and 100~TeV, and we predict borderline detectability with the Cherenkov Telescope Array (CTA). Moreover, the detection of a $\gamma$-ray counterpart might suggest larger IBH surface densities that the ones considered up to date (see Sect.~\ref{Subsec:n_sources}). On the other hand, $pp$ collisions also lead to electron-positron pair production \citep{Kelner_2006}. In turn, these particles can also contribute to the leptonic emission, but given the modest pair injection luminosity (about half of the $pp$ $\gamma$-ray luminosity), this contribution would be well below that of the primary electrons.

%====================================================================
\subsection{Luminosity scaling on the accretion rate}\label{Subsec:lambda_acc}

The energy budget of the system depends on the accretion rate of the IBH. It is then compelling to predict how the emission from the disc, the interaction structure and the extended region scales with the accretion rate, as well as on the other relevant physical quantities. The emission of an ADAF scales with the accretion rate as $L_{\rm ADAF} \propto \dot{M}^2_{\rm IBH}$ \citep{Koerding_2006, Fender_2013}. Regarding the structure, the dominant emission processes are thermal bremsstrahlung from the shocked medium and synchrotron from the shocked outflow. The former is proportional to the luminosity injected into the medium shock, that is $L_{\rm med} \propto 0.5\,\rho_{\rm med}\,v_{\rm IBH}^3\,S \propto \rho_{\rm med}\,v_{\rm IBH}^3\,z_{\rm out}^2 \propto v_{\rm IBH}\,L_{\rm out}$. In turn, the outflow mass rate is proportional to the IBH accretion rate, and the outflow luminosity can be expressed as:
\begin{equation}
    L_{\rm out} = c_{\rm eff}\,\dot{M}_{\rm IBH}\,c^2\,,
    \label{Eq:Lout_propto}
\end{equation}
where $c_{\rm eff} < 1$ is a constant that accounts for the efficiency of converting accreted rest-mass energy into outflow energy.
Then, the thermal emission from the shocked medium scales as:
\begin{equation}
    L_{\rm ff} \propto c_{\rm eff}\,v_{\rm IBH}\,\dot{M}_{\rm IBH}\,.
    \label{Eq:Lff_propto}
\end{equation}
On the other hand, the synchrotron luminosity follows the proportionality relation $L_{\rm syn} \propto L_{\rm out}\,\left(t_{\rm esc}/t_{\rm syn}\right)$. The escape timescale inside the shocked outflow is:
\begin{equation}
    t_{\rm esc} \propto \frac{z_{\rm out}}{v_1} \propto \frac{L_{\rm out}^{0.5}}{\rho_{\rm med}^{0.5}\,v_{\rm IBH}\,v_1} \propto \frac{c_{\rm eff}^{0.5}\,\dot{M}_{\rm IBH}^{0.5}}{\rho_{\rm med}^{0.5}\,v_{\rm IBH}\,v_1}\,,
    \label{tesc_propto}
\end{equation}
while assuming Eq.~(\ref{Eq:B_1}) implies:
\begin{equation}
    t_{\rm syn} \propto \frac{1}{B^2} \propto \frac{1}{\rho_{\rm med}\,v_{\rm IBH}^2}\,.
    \label{tsyn_propto}
\end{equation}
Then, from Eqs.~(\ref{Eq:Lout_propto}), (\ref{tesc_propto}) and (\ref{tsyn_propto}), we derived:
\begin{equation}
    L_{\rm syn} \propto \frac{c_{\rm eff}^{1.5}\,\dot{M}_{\rm IBH}^{1.5}\,\rho_{\rm med}^{0.5}\,v_{\rm IBH}}{v_1}\,.
\end{equation}
Regarding the core, hadronic emission dominates the spectrum, following the relation $L_{\rm pp}^{\rm core} \propto L_{\rm inj,NT}^{\rm core}\,\left(t_{\rm esc}^{\rm core}/t_{\rm pp}^{\rm core}\right) \propto L_{\rm out}\,\left(t_{\rm esc}^{\rm core}/t_{\rm pp}^{\rm core}\right)$. Diffusion dominates the escape at high energies, and assuming Bohm diffusion with coefficient $D_{\rm Bohm}$ yields $t_{\rm diff} \propto R_{\rm core}^2\,D_{\rm Bohm}^{-1}\propto R_{\rm core}^2\,B_{\rm core}$. On the other hand, the $pp$ timescale depends on the density as $t_{\rm pp} \propto \rho_{\rm core}^{-1}$. Thus, the core hadronic emission scales according to:
\begin{equation}
    L_{\rm pp}^{\rm core} \propto c_{\rm eff}\,\dot{M}_{\rm IBH}\,R_{\rm core}^2\,B_{\rm core}\,\rho_{\rm core}\,.
    \label{Lpp_propto}
\end{equation}

The mass and velocity distributions of IBHs are relatively narrow, which means that most IBHs are  concentrated around similar masses and velocities. Then, we can express the dependence on \MdotIBH as a function of the medium density considering that \MdotIBH $\propto \rho_{\rm med}$ (see Eq.~\ref{Eq:Mdot_IBH}). Finally, we remark that the accretion rate also depends on the assumed accretion parameter $\lambda_{\rm acc}$. We adopted the value of $\lambda_{\rm acc} = 0.1$ according to previous studies on mechanical feedback \citep{B-R_2020, B-R_2021}. However,  different values for this parameter, from $\lambda_{\rm acc} \ll 1$ up to $\lambda_{\rm acc} = 1$, are proposed in the literature \citep{Agol_2002,Barkov_2012,Fender_2013,Tsuna_2018,Kimura_2021,Scarcella_2021}. Thus, \MdotIBH is still quite uncertain, rendering the predicted emission and detectability of IBHs somewhat unconstrained. 

%--------------------------------------------------------------------
\subsection{Number of expected sources and acceleration of CRs}\label{Subsec:n_sources}

In order to estimate the number of IBHs similar to the case studied here within the vicinity of the Solar System, it is necessary to consider both the number of IBHs and the volume filling factor of molecular cloud cores in the region. Adopting a distance from Earth within the Galactic plane of 1~kpc, and the IBH surface densities from \cite{Tsuna_2018}, it is estimated that there are approximately $10^6$ IBHs within that region. Furthermore, the volume filling fraction of molecular clouds with densities above $10^2$~cm$^{-3}$ is $f_{\rm f}\sim 10^{-3}$ \citep[see, e.g.][and references therein]{Tsuna_2018}, yielding a number of $\sim 10^3$ IBHs inside molecular clouds within 1~kpc from Earth. The majority of these of these objects are likely to be located in environments with densities $n_{\rm med} \sim 10^2$~cm$^{-2}$, which makes them diffucult to detect (see Sect.~\ref{Subsec:MOA} below). However, adopting the distribution of volume filling fraction with density in \cite{Tsuna_2018}, and accounting for cores with $n_{\rm med}\gtrsim 3\times 10^4$~cm$^{-3}$ (i.e. allowing for a broader range of $\eta_{\rm NT}$ values of $\sim 0.1-0.3$), we obtained a core filling fraction of $f_{\rm f}^{\rm core}\sim 10^{-7}$. This subsequently yields $\sim 0.1$ IBHs inside cores within $\sim 1$~kpc from Earth, and $\sim 0.4$ IBH within $\sim 2$~kpc, which is the reference distance adopted in this work. The predicted levels of $\gamma$-ray emission would be only marginally detectable at such a distance. Consequently, an eventual detection may indicate a higher number of IBHs than expected. Conversely, radio and X-ray radiation could be detected even at distances of $\sim 10$~kpc.

We also derived optimistic estimates of the total luminosity injected into hadronic CRs above 1~PeV and 50~GeV by Galactic IBHs. We took into account that $L_{\rm out}$ is proportional to the medium density, and adopted the maximum possible acceleration efficiency that sets the maximum energy of relativistic particles as $E_{\rm max} \approx q_e\,\sqrt{4\,\left(L_{\rm out}/2\right)/c}$ (see, e.g. \cite{Barkov_2012}; not valid for electrons when synchrotron losses are important). This approach yields higher maximum energies than those derived from Eq.~\ref{Eq:Edot_acc}, and can therefore be considered as a limiting but still possible scenario. We also assumed a hard injection spectral index (i.e. $p\lesssim -2$) that implies that a significant fraction of the non-thermal energy is injected into the highest energies. Finally, for the sake of simplicity, we assumed that 10\% of $L_{\rm out}$ goes to relativistic protons/nuclei in all outflow-medium interactions, which increases the normalisation of the non-thermal particle distribution with respect to the calculations presented above. Under all these assumptions, IBHs with outflow luminosities of $L_{\rm out} \sim 10^{36}$~erg s$^{-1}$ would be capable of accelerating CRs up to $E_{\rm max} \gtrsim 1$~PeV; an example of such a distribution is shown in Fig.~\ref{fig:diste} with a green line. 

Taking a Galactic radius of 15~kpc, and extrapolating the IBH number inside dense cores within 1~kpc from Earth to the whole Galactic disc, we obtained $L_{\rm CR}(E\gtrsim 1~{\rm PeV})\sim (0.1\,{\rm IBH/kpc}^3)(15\,{\rm kpc}/1\,{\rm kpc})^2\,(0.1\,L_{\rm out})\sim 4\times 10^{36}$~erg~s$^{-1}$. On the other hand, to accelerate CRs of $\sim 50$~GeV, a luminosity of $L_{\rm out}\sim 10^{34}$~erg~s$^{-1}$ is required, meaning medium densities $\sim 100$~cm$^{-3}$. Given that in the filling factor for that density is $f_{\rm f}\sim 10^{-3}$, we obtained $L_{\rm CR}(E> 50~{\rm GeV})\sim 4\times 10^{38}$~erg~s$^{-1}$. From this, IBHs may contribute with a $\sim 0.1$~\% and $\sim 1$~\% to the Galactic CRs above 50~GeV and 1~PeV, respectively \citep{Ryan_1972,Nagano_2000,Horandel_2003}.

%====================================================================

\subsection{MOA-2011-BLG-191/OGLE-2011-BLG-0462}\label{Subsec:MOA}

We applied our model to the system MOA-2011-BLG-191/OGLE-2011-BLG-0462 to assess its broadband detectability. We assumed the system parameters reported by \cite{Sahu_2022,Sahu_2025}, i.e. $M_{\rm IBH} = 7.1$~M$_{\rm \odot}$, $v_{\rm IBH} = 51$~km\,s$^{-1}$ and a distance to the IBH of 1.58~kpc. Moreover, \cite{Sahu_2025} considered that the IBH may be located in a dense region, and then we assumed a density of $n_{\rm med} = 10^2$~cm$^{-3}$, yielding an accretion rate of $\dot{M}_{\rm IBH} \approx 2 \times 10^{12}$~g\,s$^{-1}$,  and an accreted luminosity of $L_{\rm acc} \sim 2 \times 10^{32}$~erg\,s$^{-1}$. With regard to its electromagnetic signatures, we note that MOA-2011-BLG-191/OGLE-2011-BLG-0462 was not detected in the radio catalogues TGSS \citep{TGSS}, RACS \citep{RACS}, and VLASS \citep{Lacy2020}, imposing upper limits of  29.4~mJy, 0.6~mJy and 0.24~mJy at frequencies of 150~MHz, 1.38~GHz and 3~GHz, respectively. On the other hand, the non detection in X-rays by \cite{Mereghetti_2022} imposes upper limits of $9\times 10^{-15}$~erg\,cm$^{-2}$\,s$^{-1}$ and $2\times 10^{-12}$~erg\,cm$^{-2}$\,s$^{-1}$ between 0.5--7~kev and 17--60~keV, respectively.

\begin{figure}
    \centering
    \includegraphics[width=\linewidth]{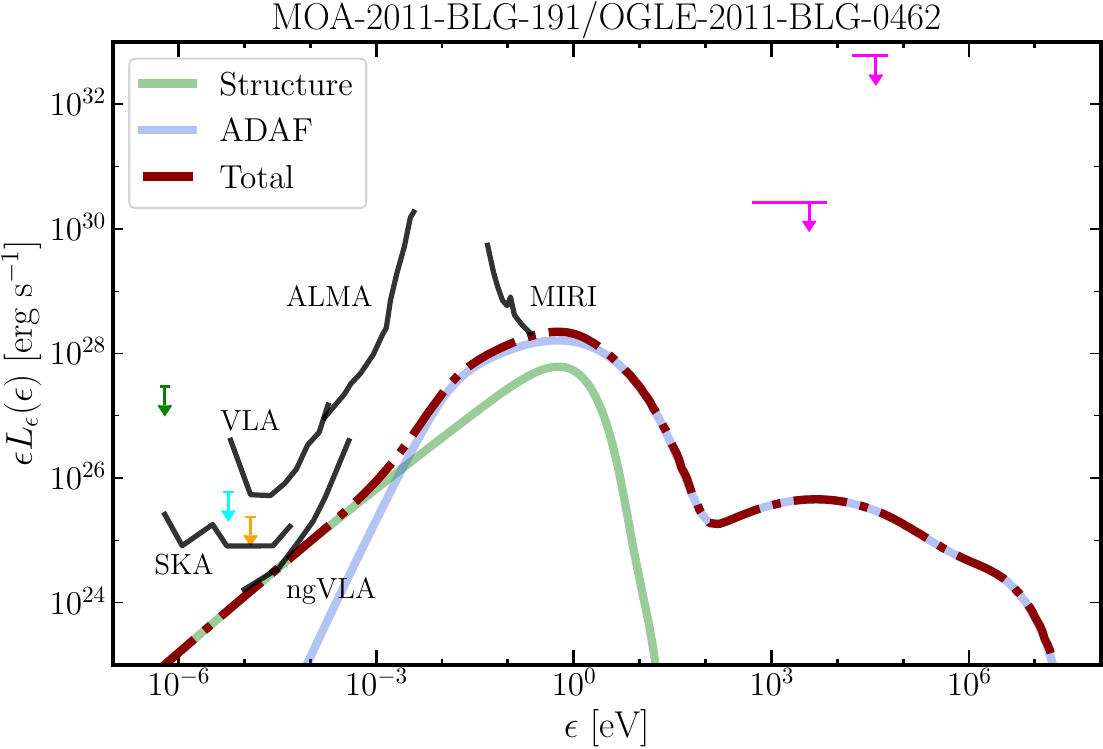}
    \caption{MOA-2011-BLG-191/OGLE-2011-BLG-0462 unabsorbed predicted spectrum. The interaction structure, accretion, and total emission are shown in light blue, light red and dark red, respectively. The 3--$\sigma$ upper limits from the TGSS, RACS and VLASS catalogues are shown with green, cyan and orange arrows, while the upper limits from \cite{Mereghetti_2022} are shown in magenta. The SKA and ngVLA sensitivity curves correspond to an integration time of 2~h \citep{Sokolowski_2022,McKinnon_2019}.}
    \label{fig:SED_MOA}
\end{figure}

In Fig.~\ref{fig:SED_MOA} we show the MOA-2011-BLG-191/OGLE-2011-BLG-0462 unabsorbed predicted spectrum. Thermal bremsstrahlung from the shocked medium dominates the emission in radio, while the accretion radiation dominates the spectrum from the millimetre up to X-rays. Regarding the outflow, assuming $v_{\rm out} = 0.5c$ yields a power of $L_{\rm out} \sim 10^{32}$~erg\,s$^{-1}$. From this total power, a luminosity of $L_{\rm inj,p} \sim 2 \times 10^{30}$~erg\,s$^{-1}$ and $L_{\rm inj,e} \sim 2 \times 10^{29}$~erg\,s$^{-1}$ is injected into protons and electrons in the deflection shocks, respectively, assuming a proton-to-electron luminosity ratio of 9. Moreover, the magnetic field results modest ($B \sim 0.1$~mG for $\eta_{\rm B} = 0.1$). Then, the resulting non-thermal emission from the shocked outflow, and the putative hadronic emission from the extended region surrounding the IBH, are negligible.

We highlight that our model is consistent with the observational upper limits. On the other hand, we predict borderline detectability in the IR with MIRI and possibly in radio with the forthcoming interferometer ngVLA\footnote{\url{https://ngvla.nrao.edu/}}, although these predictions should be taken with caution due to the large uncertainties in the system parameters \citep{Sahu_2022,Lam_2022,Lam_2023, Sahu_2025}. For instance, the detectability is contingent upon the assumption that the IBH is in a dense medium, which is not known with certainty \citep{Kimura_2025}, but the potential for detection remains promising.

%====================================================================

\subsection{Primordial black holes}\label{Subsec:PBHs}

Primordial black holes are proposed as potential constituents of the dark matter of the Universe \citep{Bird_2016,Carr_2016,Clesse_2017}. They are thought to have formed in the early Universe, during the radiation-dominated era \citep{Zeldovich_1967,Carr_1974}. If such PBHs exist, they may have affected significantly the cosmic microwave background (CMB) during the dark epoch, yielding constraints upon the mass fraction of dark matter corresponding to PBHs \citep{Piga_2022}. These constraints are dependent upon the fraction of ionising luminosity emitted by PBHs relative to the $\dot{M}_{\rm IBH}\,c^2$. The larger this fraction is, the larger the effect of each PBH on the medium, and the smaller the number of PBHs compatible with the observational constraints. We compared our results with results from \cite{Piga_2022}.

We modelled the emission of a PBH of mass $M_{\rm PBH} = 10$~M$_\odot$ moving at a velocity of $v_{\rm PBH} = 30$~km s$^{-1}$ through the Universe at redshift $z_{\rm red} = 600$. At this redshift, we considered a medium with a particle number density of $n_{\rm med} = 3\times 10^{-7}\,(1+z_{\rm red})^3~{\rm cm}^{-3} \approx 65~{\rm cm}^{-3}$ and a temperature of $T_{\rm med} = 2.73\,(1+z_{\rm red})$~K $\approx 1640$~K \citep{B-R_2020}. Furthermore, we also considered the ejection of outflows, particle acceleration at the deflection shocks and thermal emission from the shocked medium, including the effect of mechanical feedback. Under these assumptions, the main contributions to the ionising radiation are the accretion emission, the thermal emission, and potentially, the leptonic IC emission off CMB photons from the shocked outflows and the surrounding medium.

According to our model, the PBH would accrete matter at a rate of $\dot{M}_{\rm PBH} \approx 10^{13}$~g\,s$^{-1}$, yielding an accreted luminosity of $L_{\rm acc} = \dot{M}_{\rm PBH}\,c^2 \sim 10^{34}$~erg s$^{-1}$. On the one hand, the IC luminosity depends on the free parameters of the model. Assuming that 1\% of the deflection shock luminosity is reprocessed and injected into electrons yields an IC luminosity of $L_{\rm IC} \sim 3 \times 10^{29}$~erg s$^{-1}$. Regarding the accretion emission, the PBH ADAF luminosity corresponding to the accretion rate above is $L_{\rm ADAF} \sim 10^{30}$~erg s$^{-1}$. On the other hand, the shocked medium emits a luminosity of $L_{\rm ff} \sim 3 \times 10^{29}$~erg s$^{-1}$ through thermal bremsstrahlung. This means that the total ionising luminosity is $L_{\rm ion} \sim 1.6 \times 10^{30}$~erg s$^{-1}$, and the ratio of ionising-to-accretion luminosity is approximately $\sim 10^{-4}$. According to \cite{Piga_2022}, this implies that the PBH should constitute $\lesssim 10$\% of the total dark matter, assuming a monochromatic mass distribution of PBHs. In the extreme case where all the deflection shock luminosity went into non-thermal electrons, the IC luminosity would increase by a factor of 100, and the mass fraction in PBHs would be approximately of 10$^{-4}$. Therefore, our study suggests that if PBH outflows were efficient electron accelerators, their IC emission would lead to significant constraints on the PBH abundance.

%====================================================================

\section{Conclusions}\label{Sec:Conclusions}

We have made predictions on the radiation associated with IBHs, in order to address their detectability and identification in the Galaxy. In particular, we considered the development of an ADAF and the presence of outflows that interact with the surrounding medium via mechanical feedback. Subsequently, we considered the potential thermal and non-thermal radiation generated in multiple regions: the accretion structure, the region of interaction between the outflow and the medium, and the non-thermal radiation of relativistic particles escaping from the outflow. To calculate the radiation from accretion, we applied the model developed by \cite{Gutierrez_2021}. To study the radiation from the interaction region, we developed a multi-zone model, which also enabled us to study the hydrodynamics of the interaction. Finally, we employed one-zone approximations for the calculation of the core and cloud non-thermal radiation.

Our results show that a dense medium is needed to detect the electromagnetic signatures from IBHs in the Milky Way. For this reason, we modelled the passage of an IBH through the core of a molecular cloud. Our estimations suggest that one of such events could be occurring within a few kiloparsecs. When crossing these dense regions, the IBH would significantly increase its accretion rate, leading to an increase in the energy budget to fuel the radiation produced by accretion, the shocked medium and the non-thermal particles. However, molecular clouds present observational difficulties for detection in some bands because of their high emission in the millimetre and IR, and significant absorption in the optical, ultraviolet, and in soft X-rays. 

According to our work, the brightest region of the system would be the accretion structure, which would be detectable in the mid-IR and in hard X-rays with current instruments, and in $\gamma$-rays in the near future. On the other hand, the interaction region could generate multi-wavelength emission. In particular, we simulated emission maps at different frequencies, and predicted detectability in radio with instruments such as VLA and in the millimetre with ALMA if the system is at a distance of a few kiloparsecs. In addition, the interaction region could be an efficient multi-TeV accelerator. The relativistic electrons and protons would escape from the interaction structure without radiating significantly, while electrons with energies $\gtrsim 120$~GeV would cool due to synchrotron losses. Finally, the escaping particles could interact with matter, radiation and magnetic fields, both within the core and in the outer regions of the molecular cloud. This would also result in the generation of multi-wavelength emission of both hadronic and leptonic origin. In particular, we highlight the possibility of $\gamma$-rays from {\it pp} collisions in the dense medium, which could be detectable by CTA if the IBH is located at a distance of $\lesssim 2$~kpc.

We conclude that the search for electromagnetic emission from IBHs should focus on observations inside molecular clouds. In particular, the only IBH observed to date, MOA-2011-BLG-191/OGLE-2011-BLG-0462 could be detected in radio and in the IR if the IBH is located within a dense enough medium. Regarding the spectral distribution of the emission, the observed spectrum of the accretion emission of IBHs should resemble that of a microquasar in the low-hard state, but without periodic variability in the unabsorbed emission that may be associated with a binary system. Furthermore, observations in the radio and millimetre bands would be essential for identifying the outflow-medium interaction structure, and comparing the emission from this structure and that of accretion could provide information on the role of mechanical feedback and particle acceleration. The presence of a $\gamma$-ray counterpart would facilitate the identification of the presence of an IBH, and would imply that the IBH density could be higher than the one considered. Finally, we estimate that IBHs could contribute to $\sim 0.1$\% and $\sim 1$\% of the total Galactic CRs power above 50~GeV and 1~PeV, respectively. Also, a collateral conclusion of our work, once we have estimated the radiation of different origins produced by IBHs, is that PBHs cannot be a significant component of dark matter if their outflow-medium interaction structures are efficient leptonic accelerators.

%--------------------------------------------------------------------

\begin{acknowledgements}
We thank the referee, Dr. Maxim Barkov, for his insightful and constructive reports that significantly improved the manuscript.
J.R.M. and F.L.V. acknowledge support by PIP 2021-0554 (CONICET).
VB-R acknowledges financial support from the State Agency for Research of the Spanish Ministry of Science and Innovation under grants PID2022-136828NB-C41 and CEX2024-001451-M funded by MICIU/AEI/10.13039/501100011033/ERDF/EU, and from Departament de Recerca i Universitats of Generalitat de Catalunya through grant 2021SGR00679. V.B-R. is Correspondent Researcher of CONICET, Argentina, at the IAR. We sincerely thank Dr. E. M. Guti\'errez for generously providing us access to his numerical code, which was instrumental in computing the ADAF solutions.

\end{acknowledgements}


%-------------------------------------------------------------------

\bibliographystyle{aa} % style aa.bst
\bibliography{bibliography} % your references Yourfile.bib

%-------------------------------------------------------------------%-------------------------------------------------------------------

\begin{appendix}
\section{Alternative scenarios}\label{Appendix_discs}

\begin{figure}[t]
    \centering
    \includegraphics[width=0.49\textwidth]{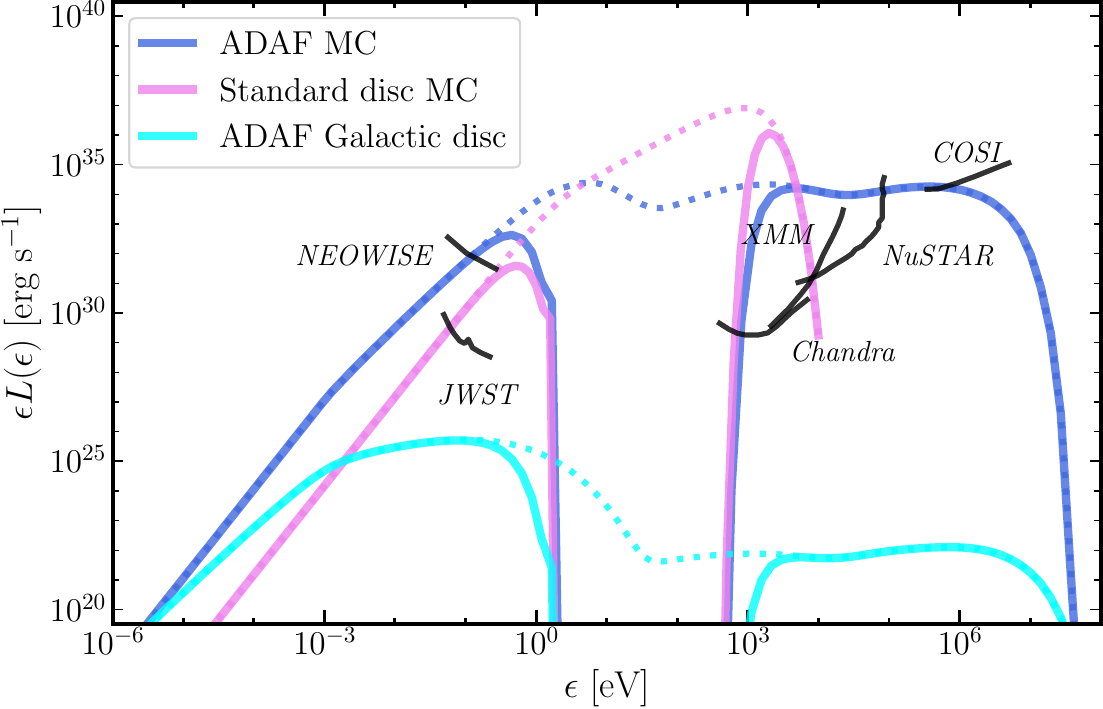}
    \caption{Comparison of the spectra emitted in three different scenarios: an ADAF within a molecular cloud core, an ADAF within a typical region of the Galactic disc, and a thin disc within a molecular cloud core. Dotted lines correspond to the intrinsic spectra, while the spectra corrected by absorption are shown in solid lines.}
   \label{fig:SED_disks_all}
\end{figure}

Here we compare the prototype scenario presented in Table~\ref{table:IBH_parameters} with two alternatives and show the results in Fig.~\ref{fig:SED_disks_all}. On the one hand, we show the resulting ADAF spectrum for an IBH in a typical density region of the Galactic disc. In this case, we considered a density of $n_{\rm med} = 1$~cm$^{-3}$, which yields an accretion rate 10$^5$ times smaller. Consequently, the resulting spectrum is negligible. This is in accordance with the non-detection of electromagnetic signatures from IBHs, given that the vast majority of IBHs should be located in similar regions of the Galactic disc. On the other hand, we considered the hypothetical formation of a thin accretion disc. Although IBHs are expected to develop an ADAF due to the accretion rate being significantly below the Eddington limit, it is expected the presence of a dense bow shock in front of the fast-moving IBH that may break up due to Rayleigh-Taylor instabilities \cite[e.g.][]{P&R_2013,B-R_2020}. As a result of this process, an excess of material may fall towards the IBH, causing a burst in the accretion, and a standard disc could temporarily develop \citep{Matsumuto_2018}. Applying the \cite{Shakura_Sunyaev_1973} model, we obtained that this type of disc also features a detectable X-ray spectrum for an accretion rate of $\dot{M}_{\rm IBH} \approx 2 \times 10^{16}~$g s$^{-1}$.

\end{appendix}
%-------------------------------------------------------------------%-------------------------------------------------------------------

\end{document}